\documentclass[10pt,letterpaper,english]{article}
\usepackage{mathptmx}
\usepackage[T1]{fontenc}
\usepackage[utf8]{inputenc}
\usepackage{amsmath,amssymb,mathrsfs}
\usepackage{graphicx}
\usepackage{amsthm}
\theoremstyle{definition}

\numberwithin{equation}{section}
\usepackage{setspace}
\usepackage{needspace}
\usepackage{amsthm} 
\usepackage{babel}
\usepackage{xcolor}
\definecolor{ReferenceGreen}{RGB}{0,128,0}
\definecolor{ReferenceRed}{RGB}{204,0,0}
\usepackage{hyperref}
\hypersetup{
  colorlinks=true,
  linkcolor=ReferenceRed,
  citecolor=ReferenceGreen,
  urlcolor=ReferenceGreen,
  filecolor=ReferenceGreen,
  pdfborder={0 0 0},
  bookmarksnumbered=true,
  bookmarksopen=false,
  linktoc=all,
  pdftitle={Simplified Parisi Ansatz III: Constructive Cavity Method and Ultrametricity},
  pdfauthor={Simone Franchini},
  pdfsubject={Third paper in the Simplified Parisi Ansatz series},
  pdfkeywords={Sherrington-Kirkpatrick model, Gibbs weights, Constructive Cavity Method, REM universality, magnetization shells, ultrametricity}
}

\usepackage{amsmath}

\AtBeginDocument{
  \setlength{\abovedisplayskip}{12pt}
  \setlength{\belowdisplayskip}{12pt}
  \setlength{\abovedisplayshortskip}{12pt}
  \setlength{\belowdisplayshortskip}{12pt}
}

\title{Simplified Parisi Ansatz III:\\
Constructive Cavity Method and Ultrametricity}
\author{Simone Franchini}
\date{}

\begin{document}
\maketitle
\begin{center}
\textit{\small CNR-ISTC, Via Gian Domenico Romagnosi 18, 00196 Rome, Italy}
\end{center}
\vspace{14pt}

\begin{abstract}
\setstretch{1.1}
We present a physical explaination of the Constructive Cavity Method (CCM) for the Sherrington--Kirkpatrick model, which identifies the decomposition of the cavity variables into layers with the random overlap structure of Aizenman--Sims--Starr. We describe the low-temperature excitations through a dilute gas of defects relative to the ground state. The sharing of these defects among the descendants and the renewal of the fluctuations along new branches produce the GREM structure and the ultrametricity of the overlaps. This mechanism is developed first with a finite number of defects and then extended to exponential populations of states. At fixed kernel, the Gaussian covariances of the fresh fields connect the complete cavity variables, including their common components, to the same hierarchy that determines the Gibbs masses. Finally, we derive the operational form of the CCM by representing the physical overlaps through suitable block dimensions, preserving the state masses and the joint law of the cavity fields and thus yielding the same Parisi functional. We also correct some ambiguities of earlier formulations.
\end{abstract}

\newpage

\clearpage
\begingroup
\setstretch{1}
\fontsize{10}{12}\selectfont
\setcounter{tocdepth}{2}
\makeatletter
\renewcommand*{\l@section}[2]{%
  \addvspace{0.2pt}%
  \@dottedtocline{1}{0em}{1.5em}{\bfseries #1}{\bfseries #2}}
\makeatother
\pdfbookmark[1]{Contents}{general-contents}
\tableofcontents
\clearpage
\endgroup

\newpage

\section{Introduction}
\label{sec:introduction}

\begingroup
\emergencystretch=1em
In this third paper in the \emph{A simplified Parisi Ansatz} series we develop a detailed  justification of the Constructive Cavity Method (CCM)~\cite{SimplifiedParisi,RSBWithoutReplicas} for the Sherrington--Kirkpatrick (SK) model~\cite{SherringtonKirkpatrick}. In its minimal operational formulation, the CCM identifies the Blanket decomposition~\cite{SimplifiedParisi,RSBWithoutReplicas,BlanketTheory,Multispin} of the cavity variables (constructive cavity variables) with the \emph{Random Overlap Structure} (ROSt) of Aizenman--Sims--Starr~\cite{AizenmanSimsStarrOriginal} described in Ref.~\cite[Sec.~2.1]{AizenmanSimsStarr}. The main objective of the present work is to explain how this identification is motivated by the energy fluctuations of the reservoir. On this basis, we address two further questions: the origin of ultrametricity in the kernel representation and the meaning of the product of Poisson point processes (PPPs) used in previous works~\cite{ConstructiveCavity}.
\par\endgroup

The first paper in the series, Ref.~\cite{SimplifiedParisi}, introduces the decomposition into layers and applied it to the cavity variables of SK, obtaining a recursion for the Parisi functional. Reference~\cite{RSBWithoutReplicas} introduces the kernel representation and developed the interface calculation through Random Energy Model (REM) universality~\cite{DerridaREM}. The second paper in the series, Ref.~\cite{REMUniversality}, further investigates this mechanism for linear energies and the role of fluctuations in magnetization selection. The Blanket theory of Ref.~\cite{BlanketTheory} then systematizes the decomposition of the interactions and the study of its scaling limits. Applications to Ising restricted and deep Boltzmann machines with Gaussian couplings~\cite{CCMBoltzmannMachines} and to the Ising perceptron~\cite{CCMIsingPerceptron} demonstrate the generality of the CCM beyond the SK model.

The physical formulation of the cavity method without replicas by Mézard, Parisi, and Virasoro derives the Parisi recursion from the ultrametric organization of the states and the exponential statistics of their free energies~\cite{MezardParisiVirasoroCavity}. Here, the Blanket decomposition and the defect gas provide a microscopic motivation for the hierarchical description used in the cavity formula.

The Blanket decomposition separates the interactions within each layer, the \emph{cores}, from those with the preceding layers, the \emph{interfaces}. By controlling the contribution of the cores, we show that the interface-only (IO) model yields the free energy per spin of SK when the thermodynamic limit is followed by refinement of the partition. The control is uniform in temperature and also preserves the minimum energy per spin.

The physical justification is developed starting from low-temperature excitations around the ground state of the IO model. Rectification with respect to a single global ground state represents this reference with all spins positive and describes the excitations through flipped sites, or defects. Thermal fluctuations favor low-cost excitations; in the nearly aligned sector, their description takes the form of a dilute gas of defects~\cite[Lemma~12]{RSBWithoutReplicas},~\cite[Secs.~3.5 and~6.2]{REMUniversality}. Describing the excitations through sets of flipped spins also allows a comparison with droplet theory~\cite{FisherHuseDroplet}, which we make precise through the decomposition of their energy cost over the interaction blocks. Fixing the number of defects in each layer specifies the magnetization shells, whose product defines a fiber. Disintegration of the interface over these shells separates the energy center from the fluctuations described by the J field~\cite{REMUniversality}. The weight of the fiber retains both contributions and the entropy of the defect arrangements.

Dilution makes the mechanism of REM universality visible~\cite{MertensRandomCosts,BaukeFranzMertens,BenArousGayrardKuptsov,BovierKurkovaLocalREM,BorgsChayesMertensNairI,BorgsChayesMertensNairII,REMUniversality,REMLinear,REMLinearLetter}: at fixed field, the different defect arrangements sample its coefficients, producing new fluctuations on the unshared supports. We first examine a finite number of defects and nodes. In this case, the interactions between defects become negligible in the thermodynamic limit, and the new arrangements can be disjoint across the entire strip of the kernel, including between nodes with different parents. Descendant states share the spins, defects, and energy contributions of their ancestors; after the branches separate, new nodes add new fluctuations. The last common prefix thus determines both the shared part of the energy and the hierarchy of distances between configurations. This construction provides the physical motivation for the structure of the Generalized Random Energy Model (GREM)~\cite{DerridaGREM}.

The passage to the extensive regime preserves the same organization over exponential populations of states~\cite{REMUniversality,REMLinear,REMLinearLetter}. We take the thermodynamic limit at fixed number of layers and fixed positive defect densities, subsequently considering the dilute regime. Controlling the relative intersections of the supports allows exponential populations to be chosen simultaneously in all strips of the kernel, while keeping the corrections to the ultrametric geometry small. The new draws are independent at fixed prefixes; their description by a common energy law for the nodes of the same level constitutes the physical GREM approximation in the extensive regime. The populations and energy parameters initially remain free, subject to the constraints of the construction.

The balance between energy and entropy is developed using the linear increments with a common law at each level and retaining the fiber energy center, while the effect of the bilinear remainder on the free energy per spin is controlled in the dilute regime. The law of the increments is determined by the population of coefficients after rectification with respect to the ground state and by sampling on the shell. The dispersion of these coefficients motivates the regularity required for counting rare energies, within the framework of REM universality for linear energies in Refs.~\cite{REMLinear,REMLinearLetter}. Comparing the rarity of the energies with the number of configurations available at each prefix determines which layers condense together and the critical temperatures of the groups~\cite{BovierKurkova,JanaRaoLDP}. In the GREM representation, the local counting of extreme energies then leads to PPPs; the slopes of the energy edge determine their exponents and, in the condensed regime, the normalized weights.

The calculation with PPPs corrects the erroneous identification of the cascade with a Cartesian product of independent REM measures in Ref.~\cite{ConstructiveCavity}. We construct the correct representation using an independent PPP per level, ancestor assignment, and global normalization of the masses. The product concerns the law of the independent processes used in the construction; the assignments reconstruct the Ruelle cascade (RPC)~\cite{RuelleCascades}, with ultrametric spin overlaps. The coagulation and fragmentation operations on Poisson--Dirichlet weights~\cite{PitmanCoagulation} allow the total, the complete masses of the prefixes, and the ancestry to be preserved jointly. Transporting the observables through the same assignments also preserves the random cavity functional in distribution. Evaluating its average corrects the step in Eqs.~(35)--(40) of Ref.~\cite{ConstructiveCavity}: the compensation of the residual factors is justified in the calculation of the mean logarithms, yielding the power means of the Parisi recursion.

The connection with the ROSt uses the same organization of the states for the Gibbs masses and for both constructive cavity variables. At fixed kernel, the fresh fields are Gaussian and their covariances are determined by the spin overlaps. Rectification with respect to the ground state preserves these overlaps and, owing to the independence of the fresh noise from the disorder of the reservoir, preserves the joint law of the fields. The amplitudes of their increments thus acquire a geometric meaning, expressed by the volume fractions and magnetizations of the layers. The calculation includes the common components: that of the linear field remains inside the hyperbolic cosine and is integrated at the end of the recursion; that of the correction field appears additively in the random functional and has zero mean. The joint representation of the masses, the overlaps, and the complete fields therefore allows the two cavity contributions to be evaluated on the same ROSt and the Parisi functional to be obtained. Minimization selects the parameters of the representation, while the subsequent refinement of the hierarchy completes the connection with the thermodynamic pressure of SK.

Subsection~\ref{subsec:operational-form-of-the-ccm} establishes the operational form of the CCM within this hierarchical representation. We choose a constructive partition whose block fractions represent the contribution common to all states and the successive increments of the physical overlap. With the same masses and ancestry, the fields built from these block dimensions have the same joint law as those of the physical representative and yield the same random cavity functional in distribution. The physical and constructive descriptions thus give the same statistical response of the reservoir in the cavity formula. Subdivision at constant mass exponent leaves both terms of the Parisi functional unchanged. Consequently, the block fractions and mass exponents used in the operational calculation parametrize the same functional obtained from the physical overlaps.

The final comparison between the Blanket measure and the Gibbs measure clarifies the use of locally normalized weights in the cavity formula of previous works~\cite{SimplifiedParisi,RSBWithoutReplicas,BlanketTheory}. The two measures are related by a tilt given by the product of the local normalizing factors, which changes the masses of the continuations and the weights of the fibers. The stepwise model of Ref.~\cite{BlanketTheory} allows the partition to be refined by jointly fixing the field values, volumes, and magnetizations of the sublayers. On each fiber specified in this way, the tilt is constant: the normalized restrictions of the two measures coincide and the common factor cancels in the conditional cavity contributions. Recomposition retains the Gibbs weights of the different fibers.

Section~\ref{sec:blanket-framework} develops the Blanket decomposition, rectification with respect to the ground state, and the description of the defect gas on the fibers. Section~\ref{sec:grem-universality} constructs the GREM representation, determines the effective levels, and clarifies the product of PPPs. Section~\ref{sec:cavity-rost} deals with the cavity formula, the identification with the ROSt, the Parisi recursion, the operational form of the CCM, and the Blanket--Gibbs tilt. Section~\ref{sec:conclusions} brings together the results and their consequences for the CCM.

\section{Blanket decomposition and rectification with respect to the ground state}
\label{sec:blanket-framework}

We decompose the SK Hamiltonian into the cores of the individual layers and the interfaces connecting them to the preceding layers~\cite{SimplifiedParisi,RSBWithoutReplicas,BlanketTheory}. Controlling the contribution of the cores shows that, after the thermodynamic limit, refinement of the partition allows the free energy and the minimum energy per spin to be calculated using the interface-only Hamiltonian. We then rectify the configurations, fields, and Gibbs measure with respect to a single global ground state of this Hamiltonian, preserving energies and overlaps. In the new coordinates, the excitations become sets of flipped spins: we calculate their complete cost, separating the isolated costs from the interactions between defects, and introduce the J field to describe the interface fluctuations~\cite{RSBWithoutReplicas,REMUniversality}. We then disintegrate the measure over the magnetization shells, whose product defines the fibers, retaining both the complete energy and the entropy of the defect arrangements in their weight. We calculate means and variances of the fields on the shells and study how the center and amplitude depend on the source prefix. Finally, we specify the selection of the physical fiber through the balance between energy and entropy and the low-temperature regime from which the hierarchical construction starts.

\subsection{Blanket decomposition and interface dominance}
\label{subsec:decomposizione-blanket-e-dominanza-dell-interfaccia}

We consider $N$ Ising spins~\cite{SherringtonKirkpatrick}, with site set and configuration space
\begin{equation}
 V=\{1,\ldots,N\},\qquad \Omega=\{-1,1\},\qquad \sigma_V\in\Omega^V.
 \label{eq:spin-space}
\end{equation}
The interaction coefficients are independent standard Gaussians for all ordered pairs of sites. We adopt the normalization of Refs.~\cite[Eq.~(2)]{SimplifiedParisi} and~\cite[Eq.~(4.6)]{RSBWithoutReplicas}:
\begin{equation}
 H(\sigma_V)=\frac1{\sqrt N}\sum_{(i,j)\in W}\sigma_iJ_{ij}\sigma_j,\qquad W=V^2.
 \label{eq:hamiltonian}
\end{equation}
The inverse temperature remains outside the Hamiltonian. The partition function and the Gibbs measure are
\begin{equation}
 Z=\sum_{\sigma_V\in\Omega^V}e^{-\beta H(\sigma_V)}.
 \label{eq:partition-function}
\end{equation}
\begin{equation}
 \mu(\sigma_V)=\frac{e^{-\beta H(\sigma_V)}}{Z}.
 \label{eq:gibbs-measure}
\end{equation}
This convention counts the two orientations of each pair separately. Compared with the usual SK Hamiltonian with a single Gaussian interaction per unordered pair, the energy field therefore has a larger amplitude by a factor of the square root of two, apart from the configuration-independent diagonal term~\cite[Eq.~(4.7)]{RSBWithoutReplicas}. We retain convention~\eqref{eq:hamiltonian} in all subsequent calculations.

Independently of the disorder, we choose an ordered partition of the sites into $L$ disjoint, nonempty layers:
\begin{equation}
 \mathcal V=\{V_1,\ldots,V_L\},\qquad V=\bigcup_{\ell=1}^L V_\ell,\qquad N_\ell=|V_\ell|.
 \label{eq:vertex-partition}
\end{equation}
The cumulative sets and the corresponding volume fractions are those of Refs.~\cite[Def.~4]{RSBWithoutReplicas} and~\cite[Eqs.~(38)--(44)]{BlanketTheory}:
\begin{equation}
 Q_\ell=\bigcup_{t\leq\ell}V_t,\qquad q_\ell=\frac{|Q_\ell|}{N}.
 \label{eq:vertex-filtration}
\end{equation}
\begin{equation}
 Q_0=\varnothing,\qquad q_0=0,\qquad q_L=1.
 \label{eq:filtration-endpoints}
\end{equation}
The volume fraction of an individual layer satisfies
\begin{equation}
 p_\ell=q_\ell-q_{\ell-1}=\frac{N_\ell}{N},\qquad \sum_{\ell=1}^L p_\ell=1.
 \label{eq:layer-volume-fractions}
\end{equation}
The blanket collects the interactions that enter the system when the layer is added:
\begin{equation}
 W_\ell=Q_\ell^2\setminus Q_{\ell-1}^2,\qquad |W_\ell|=N^2(q_\ell^2-q_{\ell-1}^2).
 \label{eq:blankets}
\end{equation}
Each ordered pair belongs to a single blanket. This gives the exact decomposition of Lemma~9 of Ref.~\cite{RSBWithoutReplicas},
\begin{equation}
 H_\ell(\sigma_{Q_\ell})=\frac1{\sqrt N}\sum_{(i,j)\in W_\ell}\sigma_iJ_{ij}\sigma_j,
 \label{eq:layer-hamiltonian}
\end{equation}
\begin{equation}
 H(\sigma_V)=\sum_{\ell=1}^L H_\ell(\sigma_{Q_\ell}).
 \label{eq:layer-decomposition}
\end{equation}
The blanket contains three disjoint sets of interactions, where juxtaposition of sets denotes the Cartesian product:
\begin{equation}
 W_\ell=V_\ell^2\cup(V_\ell Q_{\ell-1})\cup(Q_{\ell-1}V_\ell).
 \label{eq:core-interface-sets}
\end{equation}
The first is the core, consisting of the interactions within the layer; the other two form the interface with the preceding layers. Denoting the interface contribution in the normalization of the full system by a bar, we have
\begin{equation}
 \bar H_\ell(\sigma_{Q_\ell})=\frac1{\sqrt N}\sum_{i\in V_\ell}\sum_{j\in Q_{\ell-1}}\sigma_i(J_{ij}+J_{ji})\sigma_j.
 \label{eq:interface-hamiltonian}
\end{equation}
The same definition~\eqref{eq:hamiltonian}, applied to an individual layer with its own normalization, therefore gives
\begin{equation}
 H_\ell(\sigma_{Q_\ell})=\sqrt{p_\ell}\,H(\sigma_{V_\ell})+\bar H_\ell(\sigma_{Q_\ell}).
 \label{eq:core-interface-decomposition}
\end{equation}
The factor in front of the core converts the square root of the layer volume into the square root of the total volume.

For the layers after the first, we introduce the field of Ref.~\cite[Eq.~(21)]{SimplifiedParisi}:
\begin{equation}
 h_i(\sigma_{Q_{\ell-1}})=\frac1{\sqrt{|Q_{\ell-1}|}}\sum_{j\in Q_{\ell-1}}(J_{ij}+J_{ji})\sigma_j,\qquad i\in V_\ell.
 \label{eq:interface-field}
\end{equation}
The subscript indicates the site on which the field acts; the argument indicates the spins that generate it. For a source configuration fixed independently of the new block of disorder, the components are independent Gaussians with variance two. The sum of the two orientations is included in the field:
\begin{equation}
 \bar H_\ell(\sigma_{Q_\ell})=\sqrt{q_{\ell-1}}\sum_{i\in V_\ell}\sigma_i h_i(\sigma_{Q_{\ell-1}}).
 \label{eq:interface-field-form}
\end{equation}
The first layer has no interface:
\begin{equation}
 \bar H_1(\sigma_{Q_1})=0.
 \label{eq:first-layer-interface}
\end{equation}

Interface dominance in the limit of a fine partition is the principle discussed in Refs.~\cite{SimplifiedParisi},~\cite[Lemma~10]{RSBWithoutReplicas}, and~\cite[Sec.~3]{BlanketTheory}. To specify its thermodynamic content, we define
\begin{equation}
 \bar H(\sigma_V)=\sum_{\ell=1}^L\bar H_\ell(\sigma_{Q_\ell}).
 \label{eq:io-hamiltonian}
\end{equation}
We first take the thermodynamic limit with fixed $L$ and positive volume fractions, then refine the partition by making the relative volumes of the layers uniformly small. At fixed configuration, the energy of a core has variance equal to its number of spins. The Gaussian tail, compared with the exponential number of configurations, gives the uniform bound for almost every realization of the disorder
\begin{equation}
 \max_{\sigma_{V_\ell}\in\Omega^{V_\ell}}|H(\sigma_{V_\ell})|=O(N_\ell).
 \label{eq:core-extensivity}
\end{equation}
The constant can be chosen independently of the layer and of subsequent refinement. From~\eqref{eq:core-interface-decomposition} and~\eqref{eq:io-hamiltonian} it follows that
\begin{equation}
 \frac1N\max_{\sigma_V\in\Omega^V}|H(\sigma_V)-\bar H(\sigma_V)|=O\!\left(\sum_{\ell=1}^L p_\ell^{3/2}\right).
 \label{eq:core-vanishing}
\end{equation}
The bound vanishes under refinement, since
\begin{equation}
 \sum_{\ell=1}^L p_\ell^{3/2}\leq\sqrt{\max_{\ell\leq L}p_\ell}\longrightarrow0.
 \label{eq:partition-refinement}
\end{equation}

We introduce the partition function of the interface-only Hamiltonian:
\begin{equation}
 \bar Z=\sum_{\sigma_V\in\Omega^V}e^{-\beta\bar H(\sigma_V)}.
 \label{eq:io-partition-function}
\end{equation}
The associated Gibbs measure uses global normalization:
\begin{equation}
 \bar\mu(\sigma_V)=\frac{e^{-\beta\bar H(\sigma_V)}}{\bar Z}.
 \label{eq:io-gibbs-measure}
\end{equation}
Comparing the Boltzmann weights configuration by configuration gives
\begin{equation}
 \frac{|\log Z-\log\bar Z|}{\beta N}\leq\frac1N\max_{\sigma_V\in\Omega^V}|H(\sigma_V)-\bar H(\sigma_V)|.
 \label{eq:interface-free-energy}
\end{equation}
The bound~\eqref{eq:core-vanishing}, together with the limit~\eqref{eq:partition-refinement}, shows that the interface-only model yields the free energy per spin of SK in the stated order of limits. Since the right-hand side of~\eqref{eq:interface-free-energy} does not depend on temperature, the same bound holds for the minimum energy per spin. The subsequent analysis concerns this complete Hamiltonian and its Gibbs measure. The ground state used for rectification will be one of its global minima.

\subsection{Rectification with respect to the ground state}
\label{subsec:rettificazione-sul-ground-state}

Fixing the realization of the interactions, we choose a ground state of the entire inter\-face-only Hamiltonian:
\begin{equation}
 \bar H(\omega_V)=\min_{\sigma_V\in\Omega^V}\bar H(\sigma_V).
 \label{eq:ground-state}
\end{equation}
The reference remains the same for all layers and all configurations. We adopt the Hadamard product and the rectification notation of Refs.~\cite[Eqs.~(4.29)--(4.33)]{RSBWithoutReplicas} and \cite[Eqs.~(10), (27)]{REMUniversality}, applying them to the global reference in~\eqref{eq:ground-state}:
\begin{equation}
 \sigma_V^*=\sigma_V\circ\omega_V,\qquad \sigma_V=\sigma_V^*\circ\omega_V.
 \label{eq:ground-state-spin-map}
\end{equation}
The transformation is its own inverse. The chosen ground state is mapped to the all-positive configuration, denoted using the notation of Ref.~\cite[Eqs.~(57), (61)]{REMUniversality}:
\begin{equation}
 \omega_V\circ\omega_V=1_V.
 \label{eq:rectified-ground-state}
\end{equation}
The asterisk indicates calibration with respect to the ground state. For a function of the configurations, we define
\begin{equation}
 f^*(\sigma_V^*)=f(\sigma_V^*\circ\omega_V).
 \label{eq:ground-state-function-map}
\end{equation}
In particular, the transformed Hamiltonian preserves the energy value of the corresponding configuration:
\begin{equation}
 \bar H^*(\sigma_V^*)=\bar H(\sigma_V^*\circ\omega_V).
 \label{eq:rectified-hamiltonian}
\end{equation}
The action of rectification on the interaction coefficients is
\begin{equation}
 J_{ij}^*=\omega_iJ_{ij}\omega_j.
 \label{eq:ground-state-couplings}
\end{equation}
In the interface field, both the source spins and the spin at the receiving site are transformed. Equation~\eqref{eq:interface-field} becomes
\begin{equation}
 h_i^*(\sigma_{Q_{\ell-1}}^*)=\omega_i h_i(\sigma_{Q_{\ell-1}}^*\circ\omega_{Q_{\ell-1}}),\qquad i\in V_\ell,
 \label{eq:rectified-field-map}
\end{equation}
that is,
\begin{equation}
 h_i^*(\sigma_{Q_{\ell-1}}^*)=\frac1{\sqrt{|Q_{\ell-1}|}}\sum_{j\in Q_{\ell-1}}(J_{ij}^*+J_{ji}^*)\sigma_j^*.
 \label{eq:rectified-interface-field}
\end{equation}
The rectified field retains its own sign. The layer energy and its sum remain
\begin{equation}
 \bar H_\ell^*(\sigma_{Q_\ell}^*)=\sqrt{q_{\ell-1}}\sum_{i\in V_\ell}\sigma_i^*h_i^*(\sigma_{Q_{\ell-1}}^*).
 \label{eq:rectified-interface-energy}
\end{equation}
\begin{equation}
 \bar H^*(\sigma_V^*)=\sum_{\ell=1}^L\bar H_\ell^*(\sigma_{Q_\ell}^*).
 \label{eq:rectified-energy-sum}
\end{equation}
The first term is zero, as in~\eqref{eq:first-layer-interface}. By construction, the all-positive configuration minimizes the complete sum:
\begin{equation}
 \bar H^*(1_V)=\bar H(\omega_V)=\min_{\sigma_V^*\in\Omega^V}\bar H^*(\sigma_V^*).
 \label{eq:rectified-ground-state-minimum}
\end{equation}

The bijectivity of~\eqref{eq:ground-state-spin-map} preserves the partition function and transports the complete Gibbs measure:
\begin{equation}
 \bar Z=\sum_{\sigma_V^*\in\Omega^V}e^{-\beta\bar H^*(\sigma_V^*)}.
 \label{eq:rectified-partition-function}
\end{equation}
\begin{equation}
 \bar\mu^*(\sigma_V^*)=\bar\mu(\sigma_V^*\circ\omega_V)=\frac{e^{-\beta\bar H^*(\sigma_V^*)}}{\bar Z}.
 \label{eq:rectified-gibbs-measure}
\end{equation}
Each observable is transported together with the weight:
\begin{equation}
 \sum_{\sigma_V}\bar\mu(\sigma_V)f(\sigma_V)=\sum_{\sigma_V^*}\bar\mu^*(\sigma_V^*)f^*(\sigma_V^*).
 \label{eq:rectified-observable-average}
\end{equation}
These identities hold for every realization, before taking limits or disorder averages. At zero external field, global spin inversion preserves the energy: the complete sum retains both sectors associated with the ground state and its opposite.

Since the reference is common, the product of two spins at the same site remains unchanged. Consequently, for any set of sites and in particular on every prefix,
\begin{equation}
 \frac1{|Q_\ell|}\sum_{i\in Q_\ell}\sigma_i^*\tau_i^*=\frac1{|Q_\ell|}\sum_{i\in Q_\ell}\sigma_i\tau_i.
 \label{eq:rectified-overlap-invariance}
\end{equation}
Two configurations share a prefix in the physical coordinates if and only if they share it in the rectified coordinates. The transformation also puts all continuations of that prefix into correspondence and preserves its complete Gibbs mass. This property allows the prefix construction of the kernel and the triangular map used in the representation as a product of PPPs to be retained.

Calibration preserves the absolute values of the coefficients:
\begin{equation}
 |J_{ij}^*+J_{ji}^*|=|J_{ij}+J_{ji}|.
 \label{eq:calibrated-coupling-moduli}
\end{equation}
The ground state depends on the same interactions that are transformed. Global minimization therefore introduces correlations between the calibrated coefficients. The identities on the shells and the control of the amplitudes will be formulated at fixed disorder.

\subsection{Defect gas and J field}
\label{subsec:gas-di-difetti-e-j-field}

In the coordinates of~\eqref{eq:ground-state-spin-map}, flips relative to the ground state are the negative spins. We use the defect set of Ref.~\cite[Eq.~(57)]{REMUniversality}:
\begin{equation}
 X(\sigma_V^*)=\{i\in V:\sigma_i^*=-1\},\qquad X(\sigma_{V_\ell}^*)=X(\sigma_V^*)\cap V_\ell.
 \label{eq:defect-set}
\end{equation}
The lattice gas representation follows from the identity
\begin{equation}
 \sigma_i^*=1-2\mathbb{I}\!\left[i\in X(\sigma_V^*)\right].
 \label{eq:defect-spin-identity}
\end{equation}
The magnetization of the layer and its defect fraction satisfy
\begin{equation}
 M(\sigma_{V_\ell}^*)=\sum_{i\in V_\ell}\sigma_i^*=N_\ell m_\ell.
 \label{eq:rectified-magnetization}
\end{equation}
\begin{equation}
 \epsilon_\ell=\frac{|X(\sigma_{V_\ell}^*)|}{N_\ell}=\frac{1-m_\ell}{2}.
 \label{eq:defect-density}
\end{equation}
The dilute regime therefore corresponds to magnetizations close to one, with the convention of Refs.~\cite[Secs.~4.1 and~6.2]{REMUniversality} and~\cite[Lemma~12]{RSBWithoutReplicas}.

The cost of an isolated flip includes all interfaces incident on the site. From the minimality in~\eqref{eq:rectified-ground-state-minimum}, it follows that
\begin{equation}
 -\frac2{\sqrt N}\sum_{j\in V\setminus V_\ell}(J_{ij}^*+J_{ji}^*)\geq0,\qquad i\in V_\ell.
 \label{eq:single-defect-stability}
\end{equation}
For multiple defects, the isolated costs must be corrected on the bonds whose endpoints are both flipped. Substituting~\eqref{eq:defect-spin-identity} into the energy gives exactly
\begin{equation}
 \begin{aligned}
 \bar H^*(\sigma_V^*)-\bar H^*(1_V)
 ={}&-\frac2{\sqrt N}\sum_{\ell=1}^L\sum_{i\in X(\sigma_{V_\ell}^*)}\sum_{j\in V\setminus V_\ell}(J_{ij}^*+J_{ji}^*)\\
 &+\frac4{\sqrt N}\sum_{t<\ell}\sum_{i\in X(\sigma_{V_\ell}^*)}\sum_{j\in X(\sigma_{V_t}^*)}(J_{ij}^*+J_{ji}^*).
 \end{aligned}
 \label{eq:complete-defect-energy}
\end{equation}
The first term sums the nonnegative costs in~\eqref{eq:single-defect-stability}. The second restores the original contribution of the bonds with both endpoints flipped. The complete excitation energy is nonnegative, whereas the interaction between defects can have either sign. The formula also includes defects in the first layer, through their interactions with all subsequent layers.

The description through spins flipped relative to the ground state allows a comparison with droplet theory. In that theory, zero-temperature excitations are selected by minimizing the cost of flipping groups of spins at a fixed spatial scale~\cite{FisherHuseDroplet}. In the IO model, the support of the defects identifies the flipped group, and its boundary in the interaction graph consists of the bonds connecting it to its complement. The complete cost in~\eqref{eq:complete-defect-energy} can therefore be written as a sum over the boundary bonds:
\begin{equation}
 \bar H^*(\sigma_V^*)-\bar H^*(1_V)=-\frac2{\sqrt N}\sum_{\ell=1}^L\sum_{i\in V_\ell}\sum_{j\in Q_{\ell-1}}(J_{ij}^*+J_{ji}^*)\,\mathbb{I}[\sigma_i^*\sigma_j^*=-1].
 \label{eq:defect-boundary-energy}
\end{equation}
The indicator selects the bonds with only one endpoint flipped, for which the spin product changes sign. Each pair between layers appears in the interface of the layer with the larger index. The Blanket decomposition thus distributes the boundary of the excitation among its interaction blocks: each block is fixed by the partition of the sites, while the portion of the boundary it contains depends on the defect arrangement. In the comparison with SK, the size of the flipped group can be expressed through the overlap with the reference; the overlap distance plays a role analogous to the spatial scale in the droplet argument~\cite[Sec.~V]{LeDoussalMullerWieseAvalanches}.

To describe the fluctuations of a single interface, we fix its source prefix and center the field~\eqref{eq:rectified-interface-field}, following the identities of Refs.~\cite[Eqs.~(4.34)--(4.35)]{RSBWithoutReplicas} and~\cite[Eqs.~(102)--(111)]{REMUniversality}. The center and amplitude are
\begin{equation}
 \psi_\ell^*(\sigma_{Q_{\ell-1}}^*)=\frac1{N_\ell}\sum_{i\in V_\ell}h_i^*(\sigma_{Q_{\ell-1}}^*).
 \label{eq:field-center}
\end{equation}
\begin{equation}
 \varphi_i^*(\sigma_{Q_{\ell-1}}^*)=h_i^*(\sigma_{Q_{\ell-1}}^*)-\psi_\ell^*(\sigma_{Q_{\ell-1}}^*),\qquad i\in V_\ell.
 \label{eq:centered-field}
\end{equation}
\begin{equation}
 [\delta_\ell^*(\sigma_{Q_{\ell-1}}^*)]^2=\frac1{N_\ell}\sum_{i\in V_\ell}[\varphi_i^*(\sigma_{Q_{\ell-1}}^*)]^2.
 \label{eq:field-amplitude}
\end{equation}
In the following formulas, at fixed source, we omit the argument of the field. By definition,
\begin{equation}
 \sum_{i\in V_\ell}\varphi_i^*=0.
 \label{eq:centered-field-zero-sum}
\end{equation}
The centered part of the energy is therefore expressed as a sum over the defects:
\begin{equation}
 \sum_{i\in V_\ell}\sigma_i^*\varphi_i^*=-2\sum_{i\in X(\sigma_{V_\ell}^*)}\varphi_i^*.
 \label{eq:centered-defect-sum}
\end{equation}
The J field normalizes the centered components and their sum over the defect support:
\begin{equation}
 J_i^*=\frac{\varphi_i^*}{\delta_\ell^*},\qquad i\in V_\ell.
 \label{eq:j-field-components}
\end{equation}
\begin{equation}
 J^*(\sigma_{V_\ell}^*)=\frac1{\sqrt{|X(\sigma_{V_\ell}^*)|}}\sum_{i\in X(\sigma_{V_\ell}^*)}J_i^*.
 \label{eq:configuration-j-field}
\end{equation}
The single-index components have zero spatial mean and unit spatial second moment. For a spatially constant field, or in the absence of defects, the fluctuating contribution is zero.

The interface energy takes the form
\begin{equation}
 \bar H_\ell^*(\sigma_{Q_\ell}^*)=\sqrt{q_{\ell-1}}\,\psi_\ell^*M(\sigma_{V_\ell}^*)-2\sqrt{q_{\ell-1}}\,\delta_\ell^*\sqrt{|X(\sigma_{V_\ell}^*)|}\,J^*(\sigma_{V_\ell}^*).
 \label{eq:center-and-j-energy}
\end{equation}
The sign of the fluctuation follows from~\eqref{eq:centered-defect-sum}. The center is common to the configurations of the current shell at fixed source, but changes when the prefix varies. Recomposition of the interfaces retains this change, together with the fluctuations of the J field, and yields the complete energy in~\eqref{eq:complete-defect-energy}.

\subsection{Disintegration over magnetization shells}
\label{subsec:disintegrazione-sulle-shell-di-magnetizzazione}

The bijective rectification allows the sum~\eqref{eq:rectified-partition-function} to be organized according to the magnetizations of the rectified spins. We apply the disintegration of Refs.~\cite[Eqs.~(53)--(55), (114)]{REMUniversality} and~\cite[Eqs.~(9)--(14)]{BlanketTheory} to each layer. The admissible values are
\begin{equation}
 m_\ell\in\left\{-1,-1+\frac2{N_\ell},\ldots,1\right\}.
 \label{eq:admissible-magnetizations}
\end{equation}
The shell collects all configurations of the layer with the same magnetization:
\begin{equation}
 \Omega^{V_\ell}(m_\ell)=\{\sigma_{V_\ell}^*\in\Omega^{V_\ell}:M(\sigma_{V_\ell}^*)=N_\ell m_\ell\}.
 \label{eq:magnetization-shell}
\end{equation}
A magnetization profile determines the fiber
\begin{equation}
 \Omega^V(m_1,\ldots,m_L)=\prod_{\ell=1}^L\Omega^{V_\ell}(m_\ell).
 \label{eq:magnetization-fibre}
\end{equation}
The fibers are disjoint and cover the space of rectified configurations. Through~\eqref{eq:ground-state-spin-map}, they identify equally many disjoint sets of physical configurations, with the same cardinalities.

Equation~\eqref{eq:defect-density} fixes the number of defects in the shell, leaving their positions free. The counting is therefore binomial:
\begin{equation}
 |\Omega^{V_\ell}(m_\ell)|=\binom{N_\ell}{N_\ell(1-m_\ell)/2}.
 \label{eq:shell-cardinality}
\end{equation}
\begin{equation}
 |\Omega^V(m_1,\ldots,m_L)|=\prod_{\ell=1}^L|\Omega^{V_\ell}(m_\ell)|.
 \label{eq:fibre-cardinality}
\end{equation}
The uniform average on the fiber, in the notation of Ref.~\cite[Eqs.~(54)--(55)]{REMUniversality} applied to the different layers, is
\begin{equation}
 \langle f(\sigma_V^*)\rangle_{m_1,\ldots,m_L}=\frac{\displaystyle\sum_{\sigma_V^*\in\Omega^V(m_1,\ldots,m_L)}f(\sigma_V^*)}{|\Omega^V(m_1,\ldots,m_L)|}.
 \label{eq:uniform-fibre-average}
\end{equation}
This average factorizes into the uniform averages on the shells; the spins of each layer remain constrained by its magnetization. The factorization concerns the uniform counting, while the transported energy preserves the connection between the layers.

Grouping each term of~\eqref{eq:rectified-partition-function} into its fiber gives the identity
\begin{equation}
 \bar Z=\sum_{m_1,\ldots,m_L}|\Omega^V(m_1,\ldots,m_L)|\left\langle e^{-\beta\bar H^*(\sigma_V^*)}\right\rangle_{m_1,\ldots,m_L}.
 \label{eq:partition-disintegration}
\end{equation}
The sum runs over the grids~\eqref{eq:admissible-magnetizations}. The energy inside the average is the sum of all the interfaces~\eqref{eq:rectified-interface-energy}, in the coordinates rectified with respect to the ground state. The prefix-dependent centers in~\eqref{eq:center-and-j-energy} enter the average on the fiber together with the fluctuations.

The complete Gibbs weight assigned to each fiber is its contribution to~\eqref{eq:partition-disintegration}, divided by the global partition function:
\begin{equation}
 \frac{|\Omega^V(m_1,\ldots,m_L)|}{\bar Z}\left\langle e^{-\beta\bar H^*(\sigma_V^*)}\right\rangle_{m_1,\ldots,m_L}.
 \label{eq:gibbs-fibre-weight}
\end{equation}
The same decomposition holds for the numerator of every physical average in~\eqref{eq:rectified-observable-average}:
\begin{equation}
 \sum_{\sigma_V}f(\sigma_V)e^{-\beta\bar H(\sigma_V)}=\sum_{m_1,\ldots,m_L}|\Omega^V(m_1,\ldots,m_L)|\left\langle f^*(\sigma_V^*)e^{-\beta\bar H^*(\sigma_V^*)}\right\rangle_{m_1,\ldots,m_L}.
 \label{eq:observable-disintegration}
\end{equation}
Equations~\eqref{eq:gibbs-fibre-weight} and~\eqref{eq:observable-disintegration} thus retain the weight of all continuations of every prefix.

The entropy of the shell is described by the function in Ref.~\cite[Eq.~(56)]{REMUniversality}:
\begin{equation}
 \phi(m)=-\frac{1+m}{2}\log\frac{1+m}{2}-\frac{1-m}{2}\log\frac{1-m}{2}.
 \label{eq:shell-entropy}
\end{equation}
The function extends continuously to the extreme magnetizations, where it vanishes. In the thermodynamic limit with a fixed number of layers, Eqs.~\eqref{eq:shell-cardinality} and~\eqref{eq:fibre-cardinality} give
\begin{equation}
 \frac1N\log|\Omega^V(m_1,\ldots,m_L)|=\sum_{\ell=1}^L p_\ell\phi(m_\ell)+o(1).
 \label{eq:fibre-entropy}
\end{equation}
The number of profiles is the product of the cardinalities of the grids~\eqref{eq:admissible-magnetizations}, and therefore grows only as a power of the volume at fixed $L$. The thermodynamic selection of the fiber thus depends on the comparison between the entropy~\eqref{eq:fibre-entropy} and the contribution of the entire energy to the average in~\eqref{eq:partition-disintegration}.

\subsection{Coefficients of the rectified field}
\label{subsec:coefficienti-del-campo-rettificato}

Definitions~\eqref{eq:field-center}--\eqref{eq:field-amplitude} describe the fluctuations due to the arrangement of defects in the receiving layer and those produced by variation of the source. In both cases, we fix the calibrated disorder and use the uniform averages on the shells of the preceding subsection.

We first fix the prefix and hence the field acting on the current layer. The symmetry between the sites of the shell and the magnetization constraint give
\begin{equation}
 \langle\sigma_i^*\rangle_{m_\ell}=m_\ell.
 \label{eq:shell-spin-mean}
\end{equation}
For two distinct sites, expanding the square of the sum of the spins gives
\begin{equation}
 \langle\sigma_i^*\sigma_j^*\rangle_{m_\ell}=\frac{N_\ell m_\ell^2-1}{N_\ell-1},\qquad i\ne j.
 \label{eq:shell-spin-pair}
\end{equation}
Since the centered components sum to zero, the fluctuating contribution has zero mean on the shell. Its second moment is exactly
\begin{equation}
 \left\langle\left(\sum_{i\in V_\ell}\sigma_i^*\varphi_i^*\right)^2\right\rangle_{m_\ell}=(\delta_\ell^*)^2\frac{N_\ell^2}{N_\ell-1}(1-m_\ell^2).
 \label{eq:shell-centered-variance}
\end{equation}
In the normalization of the J field~\eqref{eq:configuration-j-field}, for a shell with defects and nonzero amplitude, this is equivalent to
\begin{equation}
 \left\langle[J^*(\sigma_{V_\ell}^*)]^2\right\rangle_{m_\ell}=\frac{N_\ell}{N_\ell-1}(1-\epsilon_\ell).
 \label{eq:shell-j-variance}
\end{equation}
The factor tends to one in the dilute regime. At finite density, the complement of the defect fraction remains, consistently with the amplitude correction in Eqs.~(123)--(124) of Ref.~\cite{REMUniversality}. Restoring the geometric coefficient of the interface, the variance of its energy around the center is
\begin{equation}
 \left\langle\left[\bar H_\ell^*(\sigma_{Q_\ell}^*)-\sqrt{q_{\ell-1}}N_\ell m_\ell\psi_\ell^*\right]^2\right\rangle_{m_\ell}=q_{\ell-1}(\delta_\ell^*)^2\frac{N_\ell^2}{N_\ell-1}(1-m_\ell^2).
 \label{eq:shell-interface-variance}
\end{equation}

We now consider the source. We keep the magnetizations of the preceding layers fixed and average uniformly on their fiber. The average factorizes between distinct layers; within each layer, Eqs.~\eqref{eq:shell-spin-mean} and~\eqref{eq:shell-spin-pair} hold. From the linearity of~\eqref{eq:rectified-interface-field}, it follows that
\begin{equation}
 \left\langle h_i^*(\sigma_{Q_{\ell-1}}^*)\right\rangle_{m_1,\ldots,m_{\ell-1}}=\frac1{\sqrt{|Q_{\ell-1}|}}\sum_{t<\ell}m_t\sum_{j\in V_t}(J_{ij}^*+J_{ji}^*).
 \label{eq:source-mean-field}
\end{equation}
The mean of the center is obtained by averaging this expression over the receiving sites. The second moment of the field, minus the square of its mean, is
\begin{equation}
 \begin{aligned}
 &\left\langle[h_i^*(\sigma_{Q_{\ell-1}}^*)]^2\right\rangle_{m_1,\ldots,m_{\ell-1}}-\left[\left\langle h_i^*(\sigma_{Q_{\ell-1}}^*)\right\rangle_{m_1,\ldots,m_{\ell-1}}\right]^2\\
 &\quad=\frac1{|Q_{\ell-1}|}\sum_{t<\ell}\frac{N_t(1-m_t^2)}{N_t-1}\left[\sum_{j\in V_t}(J_{ij}^*+J_{ji}^*)^2-\frac1{N_t}\left(\sum_{j\in V_t}(J_{ij}^*+J_{ji}^*)\right)^2\right].
 \end{aligned}
 \label{eq:source-field-variance}
\end{equation}
The second term in parentheses is the correction due to the magnetization constraint of the source. The calculation also retains the correlations between the fields produced by the different layers, through the square of the mean~\eqref{eq:source-mean-field}.

In the thermodynamic limit at fixed number of layers and fixed volume fractions, Eq.~\eqref{eq:calibrated-coupling-moduli} implies, for typical realizations,
\begin{equation}
 \frac1{N_\ell N_t}\sum_{i\in V_\ell}\sum_{j\in V_t}(J_{ij}^*+J_{ji}^*)^2\longrightarrow2,\qquad t<\ell.
 \label{eq:calibrated-block-square-limit}
\end{equation}
Multiplying the rows and columns of a block by signs also preserves its operator norm, which for the original Gaussian blocks is of the order of the square root of the total volume. The center is a linear form of the source spins; the spatial second moment of the field and the square of the center are quadratic forms with coefficient norm of the order of the inverse volume. The variances of these quantities on the shells are therefore of the order of the inverse volume: the magnetization constraint is handled by exchanging two spins in the same layer and using~\eqref{eq:shell-spin-pair}.

In particular, the center concentrates around its uniform mean on the source fiber:
\begin{equation}
 \psi_\ell^*(\sigma_{Q_{\ell-1}}^*)-\left\langle\psi_\ell^*(\sigma_{Q_{\ell-1}}^*)\right\rangle_{m_1,\ldots,m_{\ell-1}}\xrightarrow[N\to\infty]{\mathbb P}0.
 \label{eq:source-center-concentration}
\end{equation}
Averaging~\eqref{eq:source-field-variance} over the receiving sites, the correction involving the square of the block sum vanishes by the same norm bound. Subtracting the square of the center and using~\eqref{eq:calibrated-block-square-limit}, we obtain
\begin{equation}
 \begin{aligned}
 {}[\delta_\ell^*(\sigma_{Q_{\ell-1}}^*)]^2
 &-\frac1{N_\ell}\sum_{i\in V_\ell}\left[\left\langle h_i^*(\sigma_{Q_{\ell-1}}^*)\right\rangle_{m_1,\ldots,m_{\ell-1}}\right]^2\\
 &+\left[\left\langle\psi_\ell^*(\sigma_{Q_{\ell-1}}^*)\right\rangle_{m_1,\ldots,m_{\ell-1}}\right]^2-2\sum_{t<\ell}\frac{p_t}{q_{\ell-1}}(1-m_t^2)\xrightarrow[N\to\infty]{\mathbb P}0.
 \end{aligned}
 \label{eq:source-amplitude-concentration}
\end{equation}
The probability in these two formulas concerns the uniform source at fixed calibrated disorder. The contribution formed by the two quadratic averages is the spatial variance of the mean field~\eqref{eq:source-mean-field}; it retains the dependence on the ground state and on the realization.

When all source magnetizations coincide, the dependence takes a particularly simple form. Denoting their common value by the same magnetization, we have
\begin{equation}
 \psi_\ell^*(\sigma_{Q_{\ell-1}}^*)-m\psi_\ell^*(1_{Q_{\ell-1}})\xrightarrow[N\to\infty]{\mathbb P}0.
 \label{eq:source-equal-magnetization-center}
\end{equation}
\begin{equation}
 [\delta_\ell^*(\sigma_{Q_{\ell-1}}^*)]^2-m^2[\delta_\ell^*(1_{Q_{\ell-1}})]^2-2(1-m^2)\xrightarrow[N\to\infty]{\mathbb P}0.
 \label{eq:source-equal-magnetization-amplitude}
\end{equation}
The general formulas~\eqref{eq:source-center-concentration} and~\eqref{eq:source-amplitude-concentration} also apply to nonconstant profiles, in which the cross products of the mean fields of the different layers remain.

The deviations of the center per spin are of the order of the inverse square root of the volume, but in~\eqref{eq:center-and-j-energy} they are multiplied by the total magnetization of the layer. They can therefore contribute on the square-root-of-volume scale in the energy. We thus retain the prefix-dependent centers and their correlations with the J field in the complete sums~\eqref{eq:partition-disintegration} and~\eqref{eq:observable-disintegration}.

\subsection{Physical fiber and low-temperature limit}
\label{subsec:fibra-fisica-e-limite-di-bassa-temperatura}

The disintegration~\eqref{eq:partition-disintegration} selects the magnetization profiles through the complete contribution of the fibers. At fixed number of layers, the number of profiles is polynomial in the volume; the largest term therefore determines the pressure per spin:
\begin{equation}
 \frac1N\log\bar Z=\max_{m_1,\ldots,m_L}\left\{\sum_{\ell=1}^L p_\ell\phi(m_\ell)+\frac1N\log\left\langle e^{-\beta\bar H^*(\sigma_V^*)}\right\rangle_{m_1,\ldots,m_L}\right\}+o(1).
 \label{eq:physical-fibre-pressure}
\end{equation}
We call a fiber that realizes this maximum a physical fiber; in the presence of several dominant contributions, the measure retains their respective weights. The selection extends the balance between energy and entropy in Ref.~\cite[Eqs.~(139)--(140)]{REMUniversality} to the entire profile. The magnetizations of the different layers are determined together, because the prefix-dependent centers and the interactions between defects enter the same energy.

The reference in~\eqref{eq:ground-state} allows direct control of the zero-temperature limit. For each configuration, the ratio between its weight and that of the ground state is
\begin{equation}
 \frac{\bar\mu^*(\sigma_V^*)}{\bar\mu^*(1_V)}=\exp\!\left[-\beta\bigl(\bar H^*(\sigma_V^*)-\bar H^*(1_V)\bigr)\right].
 \label{eq:ground-state-weight-ratio}
\end{equation}
At finite volume and with at least two layers, the interface graph is connected. With Gaussian disorder, the minimum is almost surely unique up to global inversion. The ratios~\eqref{eq:ground-state-weight-ratio} therefore make the weight of positive-energy excitations vanish as the temperature tends to zero. In the two symmetric sectors, the rectified configurations align with the all-positive reference and its opposite, respectively.

The approach of the energy per spin to the minimum also admits a volume-uniform bound. The entropy of the Gibbs measure lies between zero and the logarithm of the number of configurations; comparing the free energy with the ground-state energy gives
\begin{equation}
 0\leq\left\langle\bar H^*(\sigma_V^*)\right\rangle_{\bar\mu^*}-\bar H^*(1_V)\leq\frac{N\log2}{\beta}.
 \label{eq:low-temperature-energy-bound}
\end{equation}
This estimate also holds when the thermodynamic limit precedes the temperature limit.

In the sector near the chosen ground state, the defect density is related to the magnetization by~\eqref{eq:defect-density}. The low-temperature mechanism suppresses flips with cost bounded away from zero and concentrates the fluctuations on sites and sets with small cost. For the linear problem, this mechanism is described in Ref.~\cite[Sec.~3.5]{REMUniversality}; in the full model, the relevant costs are those in~\eqref{eq:complete-defect-energy}.

In the SK model, the scaling analysis of the full-RSB equations at low temperature describes the linear pseudogap of the local-field distribution in the zero-temperature limit~\cite{PankovLowTemperature}.

With a fixed total number of defects, their interaction disappears in the thermodynamic limit. Indeed, calibration preserves the absolute value of each bond, and the maximum of the Gaussian coefficients grows only as the square root of the logarithm of the volume. It follows, uniformly over arrangements of a fixed number of defects, that
\begin{equation}
 \frac4{\sqrt N}\left|\sum_{t<\ell}\sum_{i\in X(\sigma_{V_\ell}^*)}\sum_{j\in X(\sigma_{V_t}^*)}(J_{ij}^*+J_{ji}^*)\right|=O\!\left(\sqrt{\frac{\log N}{N}}\right).
 \label{eq:finite-defect-interaction-bound}
\end{equation}
The excitations are then described, up to a vanishing error, by the sum of the costs of isolated flips. This is the starting case for the construction with shared defects and new defects on the different branches.

For a fixed positive density, however small, the number of defects is proportional to the volume and the shell retains the extensive entropy~\eqref{eq:fibre-entropy}. Estimate~\eqref{eq:finite-defect-interaction-bound} concerns a fixed number of defects. The extensive treatment in subsection~\ref{subsec:passaggio-al-regime-estensivo} includes the interaction between defects in the complete energy.

Alignment at finite volume and the energy estimate~\eqref{eq:low-temperature-energy-bound} thus identify the reference and the regime of the excitations. The following analysis first develops the REM mechanism and the hierarchy with few defects, then their extension to exponential populations. The parameters of the GREM representation remain to be determined by optimizing the Parisi functional, after its derivation in Section~\ref{sec:cavity-rost}.

\section{GREM universality of the interface}
\label{sec:grem-universality}

We study the REM mechanism at low temperature through the sampling of field coefficients on defect supports~\cite{RSBWithoutReplicas,REMUniversality}. With a finite number of defects and nodes, we construct a hierarchy in which descendants retain the defects and energies of their ancestors and add new fluctuations after the branches separate. The last common prefix thus determines the shared part of the energy and the distance between states, motivating the GREM structure and the ultrametric geometry of the kernel. We extend this construction to exponential populations, controlling the intersections of the supports and retaining all interactions in the energy. We specify the approximation of the energy laws through data common to nodes at the same level; with the populations and these laws fixed, we derive from the balance between energy and entropy the groups of layers that condense together and the exponents of their weights~\cite{BovierKurkova,JanaRaoLDP}. Finally, we correct the interpretation of the product of PPPs in Ref.~\cite{ConstructiveCavity}: we carry out the calculation at two levels and extend it to the entire hierarchy, constructing the cascade through independent processes and ancestor assignment. This representation jointly preserves in distribution the total, the complete masses, and the genealogy~\cite{PitmanCoagulation}, together with the observables needed for the cavity calculation.

\subsection{REM universality}
\label{subsec:universalita-rem}

The result of Refs.~\cite[Lemma~12]{RSBWithoutReplicas} and~\cite[Sec.~6.2--6.4]{REMUniversality} describes the fluctuations of the linear problem through the J field on the magnetization shells. The physical motivation for REM universality in the low-temperature limit is the dilution of defects: the unshared parts of their supports produce new fluctuations, while the shared parts retain common contributions. The amplitudes include the shell constraint, as in~\eqref{eq:shell-j-variance}. The Gaussian form concerns the regime in which the central approximation holds; for large fluctuations, the relevant energy law is used, as specified in Ref.~\cite[Sec.~6.4]{REMUniversality}. For linear energies with independent and identically distributed coefficients, Refs.~\cite{REMLinear,REMLinearLetter} establish, under suitable regularity conditions, REM universality on exponential populations of randomly selected configurations and the Poisson--Dirichlet limit of the Gibbs weights in the condensed regime.

We fix the disorder, the prefix, and thus all the values of the centered field. These values form a fixed population, from which the uniform average on the shell draws the defect sites without replacement. The identities of Subsection~\ref{subsec:coefficienti-del-campo-rettificato} give
\begin{equation}
 \left\langle\sum_{i\in X(\sigma_{V_\ell}^*)}\varphi_i^*\right\rangle_{m_\ell}=0.
 \label{eq:sampled-defect-mean}
\end{equation}
\begin{equation}
 \left\langle\left(\sum_{i\in X(\sigma_{V_\ell}^*)}\varphi_i^*\right)^2\right\rangle_{m_\ell}=(\delta_\ell^*)^2\frac{N_\ell^2\epsilon_\ell(1-\epsilon_\ell)}{N_\ell-1}.
 \label{eq:sampled-defect-variance}
\end{equation}
The center of the uncentered sum is the number of defects multiplied by the center of the field. At fixed prefix it is constant on the shell and contributes to the energy of the different prefixes.

We consider a finite family of configurations of the same layer, indexed by the states. We assign the sites of their defects without overlaps, choosing them uniformly. If the total number of assigned sites is fixed, this sampling differs from independent draws from the empirical population only through the possibility of drawing the same site twice. This probability is at most
\begin{equation}
 \frac{\left(\sum_\alpha|X(\sigma_{V_\ell}^{*\alpha})|\right)\left(\sum_\alpha|X(\sigma_{V_\ell}^{*\alpha})|-1\right)}{2N_\ell}\longrightarrow0.
 \label{eq:finite-defect-sampling-bound}
\end{equation}
At fixed population, the sums over the different supports are therefore jointly approximated by sums of independent samples. For a finite number of defects, the increments have the law of the sampled sums.

If the number of defects increases but their density tends to zero, the normalization of the J field instead leads to the central regime of sampling without replacement. A sufficient condition for no site to dominate the sum is
\begin{equation}
 \frac{\max_{i\in V_\ell}|\varphi_i^*|}{\delta_\ell^*\sqrt{|X(\sigma_{V_\ell}^*)|}}\longrightarrow0.
 \label{eq:defect-sampling-lindeberg}
\end{equation}
Under this condition and with an increasing number of defects, the J field tends to a centered Gaussian with variance one in the dilute regime. For a finite family of disjoint supports, the residual correlations due to sampling without replacement vanish with their volume fraction. This yields the mechanism of joint REM fluctuations.

As the prefixes vary, the field also changes. The difference is controlled directly from~\eqref{eq:rectified-interface-field}. For two prefixes that differ on a fixed number of sites, the norm of the interaction block, unchanged by the calibration, gives
\begin{equation}
 \frac1{N_\ell}\sum_{i\in V_\ell}\left[h_i^*(\sigma_{Q_{\ell-1}}^{*\alpha})-h_i^*(\sigma_{Q_{\ell-1}}^{*\beta})\right]^2=O(N^{-1}).
 \label{eq:finite-prefix-field-stability}
\end{equation}
The field population therefore remains stable in second moment in the construction with few defects. Variations of the centers enter the energy increments of the prefixes multiplied by the layer volume.

These calculations identify the renewal of fluctuations for a finite family of configurations sampled on the fiber. The following construction recombines these contributions along the prefixes of the kernel.

\subsection{GREM representation}
\label{subsec:rappresentazione-grem}

We first construct a finite family of configurations with a fixed number of defects per layer. The hierarchical arrangement is the one illustrated in Figures~5.2 and~5.3 of Ref.~\cite{RSBWithoutReplicas}: the states inherit the defects of their common ancestors and add new defects when the branches separate. We verify this construction on the complete energy, using field stability and the sampling of the preceding subsection.

\begin{figure}[p]
\centering
\includegraphics[width=\textwidth]{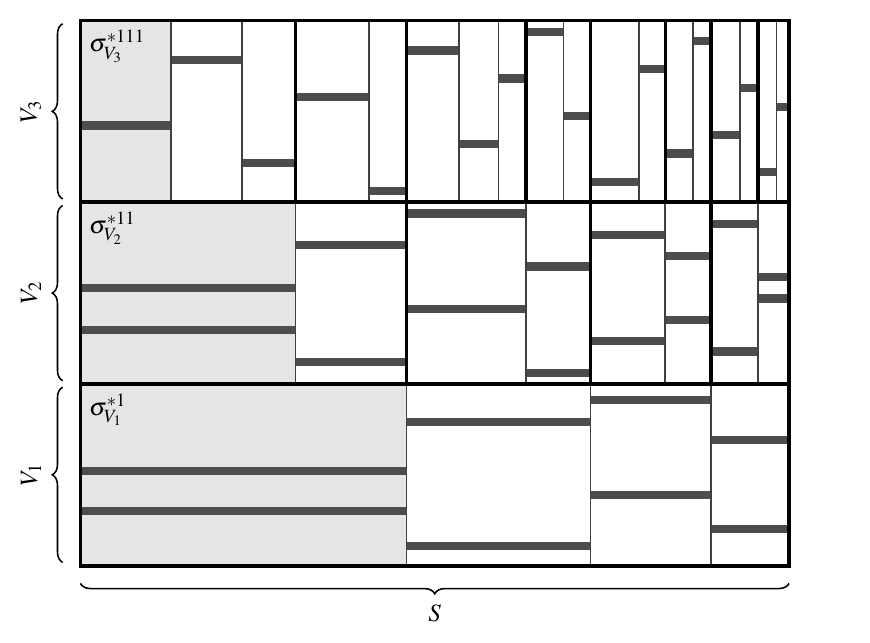}
\begingroup
\setstretch{1}
\small
\caption{Schematic representation of the GREM kernel with three layers, in the construction with finitely many defects and rectification with respect to the ground state, following the scheme of Figure~5.3 of Ref.~\cite{RSBWithoutReplicas}. The vertical coordinate runs through the sites of the layers $V_1,V_2,V_3$; the horizontal coordinate runs through the set $S$ of states, ordered by prefixes. A vertical section of the kernel identifies a complete configuration. The widths of the rectangles represent the Gibbs masses of the corresponding prefixes, including all continuations. The dark gray bands indicate spins flipped with respect to the ground state; outside the bands the spins are aligned with the reference configuration. The light gray fill highlights the layer configurations along the first branch of the hierarchy. The gray vertical boundaries separate the states of the individual layer; the black ones separate the blocks of continuations of the different preceding prefixes. Each band extends across the full width of its rectangle, making the sharing of the defect among all its descendants visible. Moving from bottom to top, each new node adds an arrangement of defects in the corresponding layer, while the descendants retain those of the ancestors in the preceding layers. The supports of distinct nodes are disjoint across the entire strip, even between different blocks. Since the number of defects is fixed in each layer, the distance between two configurations depends only on the last common prefix: this arrangement realizes the ultrametric structure described in Subsection~\ref{subsec:rappresentazione-grem}.}
\label{fig:kernel-grem-gerarchia}
\endgroup
\end{figure}

Figure~\ref{fig:kernel-grem-gerarchia} represents this arrangement of defects across three layers.

The state is identified by the sequence of indices along its branch:
\begin{equation}
 \alpha=(\alpha_1,\ldots,\alpha_L).
 \label{eq:state-path-indices}
\end{equation}
We assign to each node a support in the corresponding layer; all its descendants retain that support. The configuration on the layer therefore depends on the entire prefix:
\begin{equation}
 X(\sigma_{V_\ell}^{*\alpha})=X(\sigma_{V_\ell}^{*\alpha_1\cdots\alpha_\ell}).
 \label{eq:node-defect-support}
\end{equation}
Distinct nodes at the same level receive disjoint supports, even when they have different parents. For a finite family of nodes and a finite number of defects, this choice coincides, with probability tending to one, with independent uniform draws, by~\eqref{eq:finite-defect-sampling-bound}. The requirement therefore concerns the entire strip of the kernel. The choices in different layers can be made independently at fixed disorder.

We introduce the defects along the branch and leave the subsequent layers in the reference configuration. The composition of configurations on disjoint sets follows the notation of Ref.~\cite[Eq.~(61)]{REMUniversality}. From~\eqref{eq:complete-defect-energy} we obtain the exact increment:
\begin{equation}
 \begin{aligned}
 &\bar H^*(\sigma_{Q_\ell}^{*\alpha}\cup1_{V\setminus Q_\ell})-\bar H^*(\sigma_{Q_{\ell-1}}^{*\alpha}\cup1_{V\setminus Q_{\ell-1}})\\
 &\quad=-\frac2{\sqrt N}\sum_{i\in X(\sigma_{V_\ell}^{*\alpha})}\left[\sum_{j\in Q_{\ell-1}}(J_{ij}^*+J_{ji}^*)\sigma_j^{*\alpha}+\sum_{j\in V\setminus Q_\ell}(J_{ij}^*+J_{ji}^*)\right].
 \end{aligned}
 \label{eq:full-prefix-energy-increment}
\end{equation}
The cost of the new defects contains both the field of the prefix already chosen and that of the subsequent layers. The corrections between two defects are assigned to the layer with the larger index. For this reason, the increment of the first layer is present even if its incoming interface is zero.

The sum of the increments recombines the entire excitation:
\begin{equation}
 \begin{aligned}
 &\bar H^*(\sigma_V^{*\alpha})-\bar H^*(1_V)\\
 &\quad=\sum_{\ell=1}^L\left[\bar H^*(\sigma_{Q_\ell}^{*\alpha}\cup1_{V\setminus Q_\ell})-\bar H^*(\sigma_{Q_{\ell-1}}^{*\alpha}\cup1_{V\setminus Q_{\ell-1}})\right].
 \end{aligned}
 \label{eq:complete-prefix-telescoping}
\end{equation}
With finitely many defects,~\eqref{eq:finite-defect-interaction-bound} allows the prefix spins to be replaced by those of the reference configuration in the increment:
\begin{equation}
 \begin{aligned}
 &\bar H^*(\sigma_{Q_\ell}^{*\alpha}\cup1_{V\setminus Q_\ell})-\bar H^*(\sigma_{Q_{\ell-1}}^{*\alpha}\cup1_{V\setminus Q_{\ell-1}})\\
 &\quad=-\frac2{\sqrt N}\sum_{i\in X(\sigma_{V_\ell}^{*\alpha})}\sum_{j\in V\setminus V_\ell}(J_{ij}^*+J_{ji}^*)+o(1).
 \end{aligned}
 \label{eq:finite-node-energy-increment}
\end{equation}
The error vanishes uniformly over the finite family. The coefficients of the sum are the nonnegative complete costs of~\eqref{eq:single-defect-stability}. At fixed disorder they form a population of costs in each layer: the sampling calculation~\eqref{eq:finite-defect-sampling-bound} also applies to this population. The increments of distinct nodes are therefore jointly approximated by independent sums of costs drawn from the corresponding population. Descendants of the same node instead reuse exactly its increment.

This yields the hierarchical composition of the energy contributions~\cite{DerridaGREM}: common parts up to the last shared ancestor and new contributions after separation. The dependence on the complete prefix in~\eqref{eq:node-defect-support} comes from the arrangement of defects assigned to the node. The centers and the interactions with subsequent layers are already included in the complete costs of~\eqref{eq:full-prefix-energy-increment}. With few defects the laws of the increments remain those of the sampled sums.

The same arrangement determines the geometry of the states. We consider two branches whose last common prefix is at level $\ell$. The subsequent supports are disjoint; therefore
\begin{equation}
 \frac1N\sum_{i\in V}\sigma_i^{*\alpha}\sigma_i^{*\beta}=1-\frac2N\sum_{t>\ell}\left[|X(\sigma_{V_t}^{*\alpha})|+|X(\sigma_{V_t}^{*\beta})|\right].
 \label{eq:finite-tree-overlap}
\end{equation}
Since the number of defects is fixed by the shell of each layer, the result depends only on the last common level. The distance between states thus has the ultrametric structure of the kernel. For finitely many defects, the normalized overlaps all tend to one: the formula describes the hierarchy of small distances before this limit. The physical overlaps are identical to the rectified ones by~\eqref{eq:rectified-overlap-invariance}.

The calculation therefore describes the hierarchical mechanism and its energy increments with finitely many defects. The following subsection develops its extension to exponential populations, retaining the complete energy and leaving the populations and energy parameters of the levels to be determined.

\subsection{Passage to the extensive regime}
\label{subsec:passaggio-al-regime-estensivo}

The passage to the extensive regime takes up the mechanism already described, replacing disjointness of the supports with control of their relative intersections. We first keep the number of layers, their volume fractions, and the positive defect densities fixed. We take the thermodynamic limit and then consider the dilute regime. The shells thus have populations exponential in the volume, while the intersections can become small relative to the supports.

The construction is described starting from the terminal population. Grouping the states that share defects in the first layers, according to the scheme of Figure~\ref{fig:kernel-grem-gerarchia}, yields the nodes of the preceding levels. Adding a level means specifying further sharing of spins and defects: the terminal population can remain the same, while the preceding strips contain fewer distinct nodes. With the numbers of children of Ref.~\cite[Def.~7]{RSBWithoutReplicas}, for regular branching the count is
\begin{equation}
 n_1\leq n_1n_2\leq\cdots\leq\prod_{\ell=1}^L n_\ell=|S|.
 \label{eq:hierarchical-node-count}
\end{equation}
Each terminal state follows a single branch. Coagulation brings together configurations that share defects. The capacity of the shell imposes, in each layer,
\begin{equation}
 \prod_{t=1}^{\ell}n_t\leq\binom{N_\ell}{\epsilon_\ell N_\ell}.
 \label{eq:layer-state-capacity}
\end{equation}

Uniform sampling on the shell of Ref.~\cite[Sec.~4]{REMUniversality} allows explicit control over the entire strip. For two independent supports, the intersection has a hypergeometric law and a mean equal to the defect density multiplied by the size of a support. A bound on its tail gives, for sufficiently small densities,
\begin{equation}
 \mathbb P\!\left\{|X(\sigma_{V_\ell}^{*\alpha})\cap X(\sigma_{V_\ell}^{*\beta})|\geq N_\ell\epsilon_\ell^{3/2}\right\}\leq\left(e\sqrt{\epsilon_\ell}\right)^{N_\ell\epsilon_\ell^{3/2}}.
 \label{eq:extensive-intersection-tail}
\end{equation}
The probability here concerns the sampling of configurations at fixed disorder. Summing the bound over all pairs of nodes of the layer, including those with different parents, yields a sufficient condition for no pair to exceed this intersection, with probability tending to one:
\begin{equation}
 \limsup_{N\to\infty}\frac2{N_\ell}\log\!\left(\prod_{t=1}^{\ell}n_t\right)<\epsilon_\ell^{3/2}\log\frac1{e\sqrt{\epsilon_\ell}}.
 \label{eq:extensive-population-condition}
\end{equation}
The right-hand side is positive at sufficiently small density. For a finite number of layers, one can therefore choose exponential populations that satisfy the condition simultaneously in every strip. The constraint concerns all prefixes of the strip. The condition is sufficient and relates the exponential rate to the defect density.

The geometry of the states retains the same organization. For two branches whose last common prefix is at level $\ell$, the exact identity is
\begin{equation}
 \frac1N\sum_{i\in V}\sigma_i^{*\alpha}\sigma_i^{*\beta}=1-4\sum_{t>\ell}p_t\epsilon_t+\frac4N\sum_{t>\ell}|X(\sigma_{V_t}^{*\alpha})\cap X(\sigma_{V_t}^{*\beta})|.
 \label{eq:extensive-tree-overlap}
\end{equation}
On the event controlled by~\eqref{eq:extensive-population-condition}, the correction due to accidental overlaps satisfies
\begin{equation}
 0\leq\frac4N\sum_{t>\ell}|X(\sigma_{V_t}^{*\alpha})\cap X(\sigma_{V_t}^{*\beta})|\leq4\sum_{t>\ell}p_t\epsilon_t^{3/2}.
 \label{eq:extensive-overlap-correction}
\end{equation}
The dominant part of the distance depends only on the last common ancestor; the correction is smaller by a factor that tends to zero in the dilute regime. The structure with finitely many defects thus extends to an exponential population, retaining control over the entire strip of the kernel.

We recombine the energy exactly, including the interactions between defects and centering it on the fiber. The uniform average of the complete energy is
\begin{equation}
 \left\langle\bar H^*(\sigma_V^*)\right\rangle_{m_1,\ldots,m_L}=\frac1{\sqrt N}\sum_{t<\ell}m_tm_\ell\sum_{i\in V_\ell}\sum_{j\in V_t}(J_{ij}^*+J_{ji}^*).
 \label{eq:physical-fibre-energy-center}
\end{equation}
Subtracting this average and expanding the spin products around the shell averages, one finds
\begin{equation}
 \begin{aligned}
 &\bar H^*(\sigma_V^*)-\left\langle\bar H^*(\sigma_V^*)\right\rangle_{m_1,\ldots,m_L}\\
 &\quad=\frac1{\sqrt N}\sum_{\ell=1}^L\sum_{i\in V_\ell}(\sigma_i^*-m_\ell)\left[\sum_{j\in Q_{\ell-1}}(J_{ij}^*+J_{ji}^*)\sigma_j^*+\sum_{t>\ell}m_t\sum_{j\in V_t}(J_{ij}^*+J_{ji}^*)\right].
 \end{aligned}
 \label{eq:centered-prefix-energy-recomposition}
\end{equation}
The field in brackets contains the prefix already chosen and the average of the subsequent layers on the fiber. The formula retains all interactions exactly. The center~\eqref{eq:physical-fibre-energy-center}, constant on the fiber, remains in the comparison between profiles in~\eqref{eq:physical-fibre-pressure}.

At fixed disorder, we independently sample a uniform configuration of the shell for each node. Conditioning on all preceding levels, the fields in brackets in~\eqref{eq:centered-prefix-energy-recomposition} are fixed. The increments of the nodes at the current level are therefore functions of independent samples and their joint law factorizes exactly, even when the number of nodes is exponential. Each increment also has zero conditional mean, because the centered spin has zero mean on the shell.

This calculation identifies the shared and new parts of the complete energy. Increments at different levels have zero covariance by conditional centering; the same holds for distinct nodes at the same level. The covariance between the energies of two terminal states is therefore the sum of the variances of the increments of their common ancestors. With the same sampling at every node of a level, these variances do not depend on the node label: the energy covariance also depends only on the last common prefix. All these averages concern the configurations sampled at fixed disorder.

The geometric control~\eqref{eq:extensive-population-condition} concerns an event whose probability tends to one exponentially. Imposing this event changes the joint law of the samples by an amount that tends to zero; the exact factorization refers to independent sampling before this restriction. The two descriptions therefore have the same limit in the construction considered.

The local REM problem is the linear sum of the centered spins with the complete field in brackets in~\eqref{eq:centered-prefix-energy-recomposition}. At fixed prefix, its coefficients form a fixed population and defect sampling follows the mechanism of Ref.~\cite[Sec.~6]{REMUniversality}. Each node adds a new draw; all its descendants instead retain the same increment. The sharing of contributions up to the last common ancestor and their subsequent renewal are thus realized on the same energy that enters the measure.

The local energy law is determined by sampling supports of fixed cardinality from this population of coefficients. The regularity of its extreme energies can be motivated by simultaneously controlling the possible choices of signs for the rectification and the prefix. For each choice fixed before the disorder, the complete coefficients at different sites of the receiving layer use distinct pairs of bonds. They are therefore independent Gaussians, with variance determined by the source volumes and the magnetizations of the subsequent layers. In the nearly aligned regime this variance is positive. The total number of sign choices grows exponentially in the volume; the number of magnetization profiles instead grows as a power of the volume, at fixed partition.

For a list fixed in this way, the probability that a positive fraction of the coefficients concentrates in a very narrow interval decreases exponentially in the number of receiving sites. By narrowing the interval, the rate of this decrease can exceed that of the number of sign choices. The control then holds simultaneously for all lists and thus also for the one selected by the ground state. The same argument applies to narrow windows repeated at fixed spacing; it controls the oscillations of the energy count over finite intervals. The matrix norm, already used in Subsection~\ref{subsec:coefficienti-del-campo-rettificato}, also controls the spatial second moment and the contribution of large coefficients. A positive fraction of the sites thus retains bounded and sufficiently dispersed coefficients.

This dispersion provides the physical justification for the regularity required by the tail calculation, preserving the law of the field selected by the ground state. Condition~\eqref{eq:extensive-population-condition} leaves the sampled populations strictly below the exponential capacity of the shells. With positive densities fixed, the edge of the sampled energies therefore remains interior to the support selection problem; the following subsection determines its slope through the simultaneous constraint on energy and number of defects. The universality mechanism for non-Gaussian linear energies is developed in Refs.~\cite{REMLinear,REMLinearLetter}; here the sampling preserves the exact cardinality of the shell and regularity is motivated directly for the complete field.

The GREM description represents the new fluctuations of nodes at the same level through a common REM law, with independent realizations at different nodes~\cite{BovierKurkova}. The physical motivation is the renewal of samples already verified with finitely many defects: nodes at the same level have the same marginal sampling law, and the control~\eqref{eq:extensive-population-condition} allows the relative intersections to remain small even over an exponential population. The passage to a common law can be controlled by expanding the prefix spins around their shell averages in~\eqref{eq:centered-prefix-energy-recomposition}:
\begin{equation}
\begin{aligned}
&\bar H^*(\sigma_V^*)-\left\langle\bar H^*(\sigma_V^*)\right\rangle_{m_1,\ldots,m_L}\\
&\quad=\frac1{\sqrt N}\sum_{\ell=1}^L\sum_{i\in V_\ell}(\sigma_i^*-m_\ell)
\left[\sum_{t\ne\ell}m_t\sum_{j\in V_t}(J_{ij}^*+J_{ji}^*)\right]\\
&\qquad+\frac1{\sqrt N}\sum_{t<\ell}\sum_{i\in V_\ell}\sum_{j\in V_t}(J_{ij}^*+J_{ji}^*)(\sigma_i^*-m_\ell)(\sigma_j^*-m_t).
\end{aligned}
\label{eq:dilute-centered-energy-split}
\end{equation}
At fixed disorder and fiber profile, the coefficients of the linear term are common to all nodes of the same level. The last term contains the products of the centered fluctuations in the two layers. For a positive largest defect density not exceeding one half,
\begin{equation}
\begin{aligned}
&\frac1N\mathbb E\max_{\sigma_V^*\in\Omega^V(m_1,\ldots,m_L)}
\left|\frac1{\sqrt N}\sum_{t<\ell}\sum_{i\in V_\ell}\sum_{j\in V_t}(J_{ij}^*+J_{ji}^*)(\sigma_i^*-m_\ell)(\sigma_j^*-m_t)\right|\\
&\qquad\le C\bigl(\max_\ell\epsilon_\ell\bigr)^{3/2}\sqrt{\log\!\frac{e}{\max_\ell\epsilon_\ell}}.
\end{aligned}
\label{eq:dilute-bilinear-free-energy-bound}
\end{equation}
The constant is independent of the volume and of the chosen deterministic partition. The expectation concerns the original disorder, and the bound includes the rectification selected by its ground state.

To obtain this estimate, expand the centered spins as centered indicators of the defect supports. For supports and signs fixed before the disorder, the term with two indicators is Gaussian, with variance at most sixteen times the volume times the square of the largest defect density. The logarithm of the number of supports and sign choices on their defect sites is bounded by a constant times the volume times that density times the logarithm of its inverse. The Gaussian maximum estimate, divided by the volume, gives the right-hand side of~\eqref{eq:dilute-bilinear-free-energy-bound}. The two mixed terms and the term containing the mean densities are bounded through the norm of the symmetric matrix of interlayer bonds. Its norm divided by the square root of the volume has bounded expectation and is invariant under rectification. These contributions are absorbed in the same bound. The simultaneous maximum over signs includes the ground state selected by the disorder.

Comparing the Boltzmann sums over the same configurations, with the same multiplicities, bounds the expected absolute difference between the free energies per spin by the right-hand side of~\eqref{eq:dilute-bilinear-free-energy-bound}, uniformly in temperature. The linear energy retains the fiber center in~\eqref{eq:physical-fibre-energy-center}. The bound also compares the minimum energies per spin; for the pressure, the error is multiplied by the inverse temperature. We retain the order of limits specified above: the thermodynamic limit at fixed positive densities, followed by the dilute regime.

The common increments are identified using the normalization of~\eqref{eq:gaussian-grem-complete-energy}:
\begin{equation}
\sqrt N\bar\gamma_\ell J_\ell^{\alpha_1\cdots\alpha_\ell}
=\frac1{\sqrt N}\sum_{i\in V_\ell}(\sigma_i^{*\alpha}-m_\ell)
\left[\sum_{t\ne\ell}m_t\sum_{j\in V_t}(J_{ij}^*+J_{ji}^*)\right].
\label{eq:dilute-common-law-increment}
\end{equation}
The variance under shell sampling is absorbed in the amplitude. Independent samples at distinct nodes therefore produce increments with a common law at each level, while descendants reuse those of their ancestors. This separation supplies the common law used in the GREM description. The uniform bound~\eqref{eq:dilute-bilinear-free-energy-bound} controls the corresponding free-energy error in the dilute regime. The local laws are the full shell-sampling laws of the linear increments~\eqref{eq:dilute-common-law-increment}.

A precedent for the connection between energy selection and hierarchical organization is provided by nonhierarchical versions of the GREM: under irreducibility conditions, competition between energy and entropy produces an ultrametric structure of the relevant configurations~\cite{BolthausenKistlerUltrametricity}.

We therefore consider an extensive GREM representation with populations and energy parameters that are still free, subject to the constraints of the construction. The following subsection determines the balance between energies and entropies and the effective levels for a generic choice of these parameters. The parameters for SK will be chosen by optimizing the Parisi functional, after its derivation in Subsection~\ref{subsec:contributi-di-cavita-e-ricorsione-di-parisi}. The finite number of levels describes an approximation: the final comparison with the free energy of the model requires subsequent refinement of the hierarchy.

\subsection{Entropies, slopes, and effective levels}
\label{subsec:entropie-pendenze-e-livelli-effettivi}

The populations and energy laws of the GREM representation determine which layers condense separately and which form a single effective level. We compare the number of available choices with the rarity of the selected energies, using the full law of the increments. The calculation develops the REM balance of Refs.~\cite[Sec.~6.4--6.5]{REMUniversality} and~\cite{JanaRaoLDP}, determines the slopes of the groups, and derives their extreme weights. We keep the fiber profile, the populations, and the energy laws of the representation fixed.

The number of children of each node determines the entropy added by the layer. Up to a given prefix, the available entropy per spin is
\begin{equation}
 \lim_{N\to\infty}\frac1N\log\!\left(\prod_{t=1}^{\ell}n_t\right)=\sum_{t=1}^{\ell}\lim_{N\to\infty}\frac{\log n_t}{N}.
 \label{eq:grem-prefix-counting-entropy}
\end{equation}
The populations satisfy the capacity and dilution constraints of Subsection~\ref{subsec:passaggio-al-regime-estensivo}. The energy amplitudes of Ref.~\cite[Eqs.~(63), (85)]{SimplifiedParisi} allow the complete increments to be written in the normalized form
\begin{equation}
 \bar H^*(\sigma_V^*)-\left\langle\bar H^*(\sigma_V^*)\right\rangle_{m_1,\ldots,m_L}\overset{d}{\simeq}\sqrt N\sum_{\ell=1}^L\bar\gamma_\ell J_\ell^{\alpha_1\cdots\alpha_\ell}.
 \label{eq:gaussian-grem-complete-energy}
\end{equation}
The node variables have mean zero and variance one; their full law depends on the volume and is that of the normalized increment adopted for the level. In the GREM representation, distinct nodes receive independent realizations, while descendants reuse those of their common ancestors. The amplitudes and laws are those of the common linear increments~\eqref{eq:dilute-common-law-increment}. Their use in the free-energy calculation is controlled by~\eqref{eq:dilute-bilinear-free-energy-bound}. The rate function and the REM centering below refer to these same laws, with the disorder and fiber profile fixed. The centering is common to the nodes of a level; the fiber center remains in the energy balance and the random continuation totals enter the complete prefix masses.

We use the rate function of Ref.~\cite{JanaRaoLDP}, with the level subscript and the calibration asterisk. The argument is the energy of the level per spin of the complete system. The exponential transform of the energy law determines the cost:
\begin{equation}
 I_\ell^*(e_\ell)=\sup_{s\in\mathbb R}\left\{-se_\ell-\lim_{N\to\infty}\frac1N\log\mathbb E\exp\!\left[-s\sqrt N\bar\gamma_\ell J_\ell^{\alpha_1\cdots\alpha_\ell}\right]\right\}.
 \label{eq:grem-complete-energy-rate}
\end{equation}
The average concerns only the energy variable of the level, with the profile and the common data fixed. For the local problem, the transform is evaluated on the shell, using the common linear increment of~\eqref{eq:dilute-common-law-increment}. The sum includes all supports with the prescribed number of defects. The convexity of the transform and the dispersion of the coefficients described in the preceding subsection give the lower convex branch of the cost, with a regular slope in the selected region.

To count paths with assigned energies, the cardinalities are multiplied by the probabilities of the respective energy windows. The logarithm of the count is the available entropy minus the cost of the energies. Availability must already hold at each prefix:
\begin{equation}
 \sum_{t=1}^{\ell}I_t^*(e_t)\leq\lim_{N\to\infty}\frac1N\log\!\left(\prod_{t=1}^{\ell}n_t\right),\qquad \ell\leq L.
 \label{eq:grem-prefix-energy-cost}
\end{equation}
This yields the prefix criterion of Ref.~\cite[Theorem~3.1]{JanaRaoLDP}. For a candidate group of consecutive layers from level $\ell$ to level $t$, the lower edge exhausts the available entropy:
\begin{equation}
 \sum_{r=\ell}^{t}I_r^*(e_r)=\lim_{N\to\infty}\frac1N\log\!\left(\prod_{r=\ell}^{t}n_r\right).
 \label{eq:grem-general-group-edge}
\end{equation}
Among allocations with this cost, the minimum total energy is obtained by equating the slopes:
\begin{equation}
 -\partial_{e_r}I_r^*(e_r)=\beta_c,\qquad \ell\leq r\leq t.
 \label{eq:grem-general-group-slope}
\end{equation}
A difference between the slopes would indeed allow the overall energy to be lowered by transferring part of the cost from one layer to another. The common slope is the critical inverse temperature of the group.

Starting from the first layer still to be assigned, we compare the consecutive groups that begin at that point and choose the one with the smallest critical inverse temperature; in the event of a tie, we choose the longest. On the lower convex branch, the cost increases with the slope. Each internal prefix of the chosen group has a critical slope at least as large, so at the slope of the group its cost is still compensated by the number of available configurations. We repeat the choice on the subsequent layers. The critical inverse temperatures of the groups are increasing: a decrease would make it more favorable to include that group in the preceding choice. The saturated internal constraints enter the edge count and its centering. This extends the grouping of Refs.~\cite[Sec.~3]{DerridaGardnerGREM} and~\cite[Sec.~1.3]{BovierKurkova} to the full laws.

For the effective group, the variances along a branch add and the cardinality is the product of the numbers of children. The REM rescaling of Refs.~\cite[Lemma~13]{RSBWithoutReplicas} and~\cite[Eqs.~(133)--(136)]{REMUniversality} preserves these two quantities:
\begin{equation}
 R=\frac1{\log2}\log\!\left(\prod_{r=\ell}^{t}n_r\right).
 \label{eq:grem-group-rem-size}
\end{equation}
\begin{equation}
 b=\beta\sqrt{\frac{2N}{R}\sum_{r=\ell}^{t}\bar\gamma_r^2}.
 \label{eq:grem-group-rem-temperature}
\end{equation}
The normalized energy variable of the effective REM preserves the full law of the group. The partition function is
\begin{equation}
 Z(b,R)=\sum_{\alpha\leq2^R}\exp\!\left[-bJ^\alpha\sqrt{R/2}\right].
 \label{eq:grem-group-rem-partition}
\end{equation}
In the units of the original system, the balance between energy and entropy gives
\begin{equation}
 \lim_{N\to\infty}\frac1N\log Z(b,R)=\sup_{e_\ell,\ldots,e_t}\left\{\lim_{N\to\infty}\frac1N\log\!\left(\prod_{r=\ell}^{t}n_r\right)-\sum_{r=\ell}^{t}\left[I_r^*(e_r)+\beta e_r\right]\right\}.
 \label{eq:grem-general-group-pressure}
\end{equation}
The maximum is taken over the energies available at the prefixes of the group. Constraint~\eqref{eq:grem-prefix-energy-cost} applies starting from the first layer of the group, summing costs and entropies up to the prefix considered. Below the critical inverse temperature, the maximum is interior to the populated region and the slopes coincide with the inverse of the physical temperature. At the transition, the edge~\eqref{eq:grem-general-group-edge} is reached. In the condensed regime the slope remains that of~\eqref{eq:grem-general-group-slope}.

Obtaining the extreme weights requires counting over finite energy differences. The sum over the shell is exponentially weighted by fixing the mean number of defects and the selected energy together. The cardinality constraint and the energy fluctuation are therefore treated jointly. The margin relative to the shell capacity keeps the slope that selects the edge finite and positive; the dispersion of the coefficients guarantees a positive energy variance even after fixing the number of defects. The selected configurations thus retain regular fluctuations around the imposed energy. This is the mechanism of the precise tail evaluation of Ref.~\cite[Sec.~1.1]{BovierMayer}, applied here to sampling on the shell.

For a single list, the joint fluctuations of number and energy give a prefactor of the order of the inverse volume. The exact constraint on the number of defects introduces a normalization factor of the order of the inverse square root of the volume; the prefactor of the energy probability on the shell is therefore of the order of the inverse square root of the volume. Its relative variation vanishes when the energy is shifted by a finite amount. The edge must be calculated on the actual population of coefficients, with its dependence on the disorder and fiber profile; the prefactor is then absorbed into a logarithmic correction to the centering. For groups, the same calculation retains the constraints of the internal prefixes and the common slope~\eqref{eq:grem-general-group-slope}.

We denote by $E$ the finite displacement from the edge centered in this way. The expected number of states below a fixed threshold, calculated over branches that satisfy the internal prefix constraints, grows with the slope of the edge; choosing the origin so that the expected number below zero tends to one, the limiting energy intensity is
\begin{equation}
 \beta_c e^{\beta_c E}\,dE,\qquad E\in\mathbb R.
 \label{eq:grem-extreme-energy-intensity}
\end{equation}
For independent draws, the counts of rare energies in disjoint intervals converge to Poisson counts. In the effective groups, prefix control identifies the branches available at the edge; levels that saturate the balance together are collected in the same process. Translations common to the law of a level fix the origin of the energies; energy factors that depend on the prefix remain in the complete masses. Field-dependent centering and the local PPP limit are established for the linear model in Ref.~\cite[Theorems~1.2--1.3]{REMLinear}.

In the condensed regime, the Boltzmann weight decreases faster than the count \eqref{eq:grem-extreme-energy-intensity} grows. The evaluation on the shell bounds the mean number of states in energy intervals of fixed width, uniformly in the volume and the position of the interval, by an exponential whose slope tends to the critical one. In the condensed regime this slope remains below the inverse of the physical temperature. Multiplying by the weight, the contribution of energies above the window therefore decreases with the positive difference between these two slopes. The probability of finding states far below the window tends to zero as the window is enlarged. These two controls allow passage from the process of extremes to the total and the normalized weights. The normalized and ordered weights of each local Poisson list have a Poisson--Dirichlet law in the condensed regime. The corresponding limit of the Gibbs weights of the linear model is established in Ref.~\cite[Corollary~1.4]{REMLinear} and developed in Ref.~\cite[Sec.~\emph{A 1RSB structure for the Gibbs weights}]{REMLinearLetter}.

The change from energy to weight gives the ratio between the critical slope and the inverse temperature. With the notation of Refs.~\cite[Eqs.~(7)--(8)]{SimplifiedParisi} and~\cite[Eqs.~(126)--(128)]{REMUniversality}, we assign to all layers of the same group
\begin{equation}
 \lambda_r=\min\!\left\{1,\frac{\beta_c}{\beta}\right\},\qquad \ell\leq r\leq t.
 \label{eq:grem-group-mass-exponent}
\end{equation}
The group is condensed when the inverse temperature exceeds the critical one. At an inverse temperature below the threshold, the maximum that determines the pressure is inside the populated energy region and an exponential number of contributions remains; these thermal groups correspond to exponent one and the ordinary average. At the threshold, the maximum reaches the edge and the exponent is still one. Since the critical inverse temperatures increase along the hierarchy, the condensed groups occupy the initial part and the thermal groups the final part, if present. The exponents satisfy
\begin{equation}
 0<\lambda_1\leq\lambda_2\leq\cdots\leq\lambda_L\leq1.
 \label{eq:ordered-grem-mass-exponents}
\end{equation}
Each condensed group produces a level of the cascade, with strictly increasing exponents between distinct groups. We retain the indices of the original layers by assigning the same exponent to the merged layers. The grouping sums their energy contributions; the fiber center~\eqref{eq:physical-fibre-energy-center} accompanies this sum in the complete balance~\eqref{eq:physical-fibre-pressure}. The following subsections use the effective levels identified in this way to construct the cascade and the product of PPPs.

\subsection{GREM cascade and product of PPPs at two levels}
\label{subsec:cascata-grem-e-prodotto-di-ppp-a-due-livelli}

This section and the next correct an error in a previous article on the CCM~\cite{ConstructiveCavity} concerning the possibility of using a nonultrametric product of PPPs to obtain the Parisi functional from Aizenman's formula. In the abstract, the introduction, and Eqs.~(18)--(24) of Ref.~\cite{ConstructiveCavity}, the state is represented as a Cartesian product of independent REM measures and equated with the cascade in the calculation of the Parisi functional. This identification arises from an incorrect interpretation of the triangular map satisfied by the IO Hamiltonian. In~\eqref{eq:centered-prefix-energy-recomposition}, the law of the new draws factorizes conditionally on the preceding levels; the energy increments retain their dependence on the prefixes and are shared by the descendants.

The comparison between the representations must therefore preserve masses, ancestors, and fields together. Suppressing the prefixes in Eq.~(30) of Ref.~\cite{ConstructiveCavity}, interpreted on the Cartesian product, preserves the Gaussian marginals but alters the correlations between states. In the calculation of Eqs.~(35)--(40) of the same reference, the identities in distribution of the individual marked sums are also used to cancel residual factors between the two contributions. This cancellation requires their joint identification, which the marginal identities alone do not provide.

Below we provide the construction based on the correct interpretation through one independent PPP per level, ancestor assignment, and global normalization of the masses. The product concerns the input processes; the resulting measure is the Ruelle cascade (RPC)~\cite{RuelleCascades}, with ultrametric spin overlaps. The joint transport of the observables preserves the distribution of the random functional of the cascade; its average yields the Parisi functional, as we will develop in Subsection~\ref{subsec:contributi-di-cavita-e-ricorsione-di-parisi}. Here we carry out the calculation at two levels, tracking the total, the normalized masses, and the ancestors; in the following subsection we extend it to the entire hierarchy.

The counting in Subsection~\ref{subsec:entropie-pendenze-e-livelli-effettivi} produces the Poisson lists of extreme energies and the limit of the normalized weights of the condensed groups. Passing from the energy in~\eqref{eq:grem-extreme-energy-intensity} to the Boltzmann weight, the critical slope and the exponent~\eqref{eq:grem-group-mass-exponent} give the intensity for positive masses
\begin{equation}
 \lambda_\ell x^{-1-\lambda_\ell}\,dx,\qquad x>0,\quad 0<\lambda_\ell<1.
 \label{eq:poisson-mass-intensity}
\end{equation}
The multiplicative constant of the intensity is fixed by choosing the origin of the extreme energies for the common law of the level. Energy factors that depend on the prefix enter the complete masses. For these exponents, the small points have a finite sum and the points above a positive threshold are finite in number: the total is therefore finite almost surely.

We retain the notation for the partition function and normalized weights of the REM in Refs.~\cite[Eq.~(5.25)]{RSBWithoutReplicas} and~\cite[Eqs.~(24), (136)]{REMUniversality}. The total of each list of the cascade takes the prefix configuration as its argument, even at the first level where the prefix is empty. In the representation through independent processes, $Z_\ell$ denotes the total of the single input PPP of the level. This notation will also be used for the same processes in the calculation of the cavity contributions. A point of the local process is the product of the total and the corresponding normalized weight. At two levels, the processes of the cascade are
\begin{equation}
 \{Z_1(\sigma_{Q_0}^*)\xi_1^{\alpha_1}:\alpha_1\geq1\},\qquad \{Z_2(\sigma_{Q_1}^{*\alpha_1})\xi_2^{\alpha_1\alpha_2}:\alpha_2\geq1\}.
 \label{eq:two-level-cascade-points}
\end{equation}
The first process has the exponent of the first level; each of its points receives an independent list of children, with the exponent of the second level. The order of the condensed groups is
\begin{equation}
 0<\lambda_1<\lambda_2<1.
 \label{eq:two-level-exponents}
\end{equation}
Each local list is normalized:
\begin{equation}
 \sum_{\alpha_1}\xi_1^{\alpha_1}=1,\qquad \sum_{\alpha_2}\xi_2^{\alpha_1\alpha_2}=1.
 \label{eq:two-level-local-normalizations}
\end{equation}
The complete weight of a state is obtained by multiplying the points along its branch and normalizing by the sum over all branches:
\begin{equation}
 \frac{Z_1(\sigma_{Q_0}^*)\xi_1^{\alpha_1}Z_2(\sigma_{Q_1}^{*\alpha_1})\xi_2^{\alpha_1\alpha_2}}{\displaystyle\sum_{\beta_1}Z_1(\sigma_{Q_0}^*)\xi_1^{\beta_1}Z_2(\sigma_{Q_1}^{*\beta_1})}.
 \label{eq:two-level-complete-leaf-mass}
\end{equation}
The mass of the first prefix includes all children:
\begin{equation}
 \frac{\xi_1^{\alpha_1}Z_2(\sigma_{Q_1}^{*\alpha_1})}{\displaystyle\sum_{\beta_1}\xi_1^{\beta_1}Z_2(\sigma_{Q_1}^{*\beta_1})}.
 \label{eq:two-level-complete-parent-mass}
\end{equation}
This is the global normalization of the cascade~\cite[Sec.~5.2]{AizenmanSimsStarr}; the totals of the children weight the prefixes before normalization.

We fix the first process. Multiplying all points of the list of children by the mass of the parent transforms their intensity into
\begin{equation}
 (Z_1(\sigma_{Q_0}^*)\xi_1^{\alpha_1})^{\lambda_2}\lambda_2 x^{-1-\lambda_2}\,dx.
 \label{eq:two-level-scaled-child-intensity}
\end{equation}
The superposition of the lists is therefore a single PPP. Its intensity is the sum of the intensities in~\eqref{eq:two-level-scaled-child-intensity}. We can generate it by taking a second-level process independent of the first, with points and normalization
\begin{equation}
 \{Z_2\xi_2^{\alpha_2}:\alpha_2\geq1\},\qquad \sum_{\alpha_2}\xi_2^{\alpha_2}=1,
 \label{eq:two-level-independent-second-process}
\end{equation}
and multiplying all its points by
\begin{equation}
 Z_1(\sigma_{Q_0}^*)\left[\sum_{\alpha_1}(\xi_1^{\alpha_1})^{\lambda_2}\right]^{1/\lambda_2}.
 \label{eq:two-level-common-mass-scale}
\end{equation}
The sum is finite almost surely by the ordering~\eqref{eq:two-level-exponents}. We then assign a parent to each point of the second process, independently of the other points with the first process fixed, with probability
\begin{equation}
 \frac{(\xi_1^{\alpha_1})^{\lambda_2}}{\displaystyle\sum_{\beta_1}(\xi_1^{\beta_1})^{\lambda_2}}.
 \label{eq:two-level-parent-assignment}
\end{equation}
Separating the points according to their parent produces independent lists with the intensities~\eqref{eq:two-level-scaled-child-intensity}. It therefore jointly reconstructs the masses and their genealogy~\cite[Theorem~2.2]{BolthausenSznitmanCascades}. The superposition and splitting operations are those underlying the coagulation of Poisson--Dirichlet weights~\cite[Theorem~12]{PitmanCoagulation}.

In particular, the total satisfies the identity in distribution
\begin{equation}
 \sum_{\alpha_1}Z_1(\sigma_{Q_0}^*)\xi_1^{\alpha_1}Z_2(\sigma_{Q_1}^{*\alpha_1})\stackrel d=Z_1(\sigma_{Q_0}^*)\left[\sum_{\alpha_1}(\xi_1^{\alpha_1})^{\lambda_2}\right]^{1/\lambda_2}Z_2.
 \label{eq:two-level-total-product}
\end{equation}
In the second construction, the normalized mass of the final point is its weight in the second process. The mass of a parent is the sum of the final weights of the children assigned to it. This yields the same complete masses as in~\eqref{eq:two-level-complete-leaf-mass} and~\eqref{eq:two-level-complete-parent-mass}. The probability~\eqref{eq:two-level-parent-assignment} determines the genealogy; the subsequent sum over descendants determines the physical mass of the prefix.

\subsection{Product of PPPs and equivalence in distribution with the cascade}
\label{subsec:prodotto-di-ppp-ed-equivalenza-in-distribuzione-con-la-cascata}

The superposition of the lists and the assignment of parents are iterated along the hierarchy. We first write the case in which each layer forms a distinct condensed group:
\begin{equation}
 0<\lambda_1<\cdots<\lambda_L<1,\qquad \lambda_{L+1}=1.
 \label{eq:independent-process-exponents}
\end{equation}
We take an independent PPP for each level, with the intensity~\eqref{eq:poisson-mass-intensity}, and write its points in terms of the total and normalized weights:
\begin{equation}
 \{Z_\ell\xi_\ell^{\alpha_\ell}:\alpha_\ell\geq1\},\qquad \sum_{\alpha_\ell}\xi_\ell^{\alpha_\ell}=1.
 \label{eq:independent-local-rem-processes}
\end{equation}
Independence concerns the processes at different levels. Within each process, the total and the weights are obtained from the same points and retain their joint law~\cite[Prop.~10]{PitmanYorPoissonDirichlet}. The single index identifies a node of the level; its complete address is constructed by assigning its ancestors.

The change of variable that raises the points to the power of the next exponent produces another PPP:
\begin{equation}
 \{(Z_\ell\xi_\ell^{\alpha_\ell})^{\lambda_{\ell+1}}:\alpha_\ell\geq1\},\qquad \frac{\lambda_\ell}{\lambda_{\ell+1}}x^{-1-\lambda_\ell/\lambda_{\ell+1}}\,dx.
 \label{eq:powered-input-processes}
\end{equation}
The transformation follows directly from~\eqref{eq:poisson-mass-intensity}. The ratio between the exponents is less than one, so the sum of the new points is finite. These independent processes constitute the inputs of the product representation.

With the processes fixed, we make the assignments independently across nodes and levels. Each node at the next level receives a parent with probability
\begin{equation}
 \frac{(\xi_\ell^{\alpha_\ell})^{\lambda_{\ell+1}}}{\displaystyle\sum_{\beta_\ell}(\xi_\ell^{\beta_\ell})^{\lambda_{\ell+1}}},\qquad \ell<L.
 \label{eq:multilevel-parent-assignment}
\end{equation}
Each node receives its parent only once. All its descendants inherit this choice and the ancestors already assigned to the parent. The map is thus triangular: reconstructing a prefix uses only the assignments up to that level. The composition can be verified by starting from the last children and repeating the calculation of the preceding subsection. At each step the superposition produces a PPP; rescaling fixes its intensity and parent assignment reconstructs the original lists.

In the cascade notation, the local total depends on the configuration of the preceding prefix and the local weight carries the complete prefix. The identity of the totals is written as
\begin{equation}
 \sum_{\alpha_1,\ldots,\alpha_L}\prod_{\ell=1}^L\left[Z_\ell(\sigma_{Q_{\ell-1}}^{*\alpha_1\cdots\alpha_{\ell-1}})\xi_\ell^{\alpha_1\cdots\alpha_\ell}\right]\stackrel d=\prod_{\ell=1}^L Z_\ell\left[\sum_{\alpha_\ell}(\xi_\ell^{\alpha_\ell})^{\lambda_{\ell+1}}\right]^{1/\lambda_{\ell+1}}.
 \label{eq:multilevel-total-product}
\end{equation}
At the first level, the total on the left is that of the first process. Each factor on the right is the power of the sum of the points of~\eqref{eq:powered-input-processes}. The last factor is the total of the final process.

After global normalization, the mass of the final point in the independent representation is
\begin{equation}
 \xi_L^{\alpha_L}.
 \label{eq:independent-terminal-mass}
\end{equation}
The point occupies the complete address reconstructed through~\eqref{eq:multilevel-parent-assignment}. The mass of a prefix is obtained by summing over all final points descending from it. If the indices of the preceding levels denote the ancestors assigned to each final point, this mass is
\begin{equation}
 \sum_{\substack{\beta_L:\;(\beta_1,\ldots,\beta_\ell)=(\alpha_1,\ldots,\alpha_\ell)}}\xi_L^{\beta_L}.
 \label{eq:independent-complete-prefix-mass}
\end{equation}
The total, the prefix masses, and the genealogical relationships jointly have the law of the cascade, up to the numbering of children within each prefix~\cite[Theorem~2.2]{BolthausenSznitmanCascades}. In the inverse description, we order the prefixes according to their complete masses and divide each mass among its children. The ordered vectors of complete proportions follow the fragmentation laws
\begin{equation}
 \mathrm{PD}(\lambda_\ell,-\lambda_{\ell-1}),\qquad \lambda_0=0,
 \label{eq:complete-prefix-fragmentation}
\end{equation}
according to Corollary~13 of Ref.~\cite{PitmanCoagulation}. The second parameter incorporates the weight of the descendants in the selection of the parent.

\begingroup\emergencystretch=1em The construction preserves the geometry of the kernel illustrated in Figure~\ref{fig:kernel-grem-gerarchia}. The columns associated with descendants of the same node together occupy the mass~\eqref{eq:independent-complete-prefix-mass}; their spins share the defects of the prefix. The layer configuration and any values of the observables remain associated with each node. The correspondence transports these data together with the masses. For a function of the states, the normalized sum in the independent representation takes the form\par\endgroup
\begin{equation}
 \sum_{\alpha_L}\xi_L^{\alpha_L}f^*(\sigma_V^{*\alpha_1\cdots\alpha_L}),
 \label{eq:independent-marked-average}
\end{equation}
where all ancestors belong to the same reconstructed address. Multiple observables are transported with the same assignments. This allows the two contributions of the cavity functional to be preserved jointly, as we will develop in Subsection~\ref{subsec:contributi-di-cavita-e-ricorsione-di-parisi}.

The global rectification~\eqref{eq:ground-state-spin-map} preserves the prefixes, their continuations, and the overlaps. The mass map therefore operates in the same way on the physical and rectified configurations. For the layers merged in Subsection~\ref{subsec:entropie-pendenze-e-livelli-effettivi}, the calculation applies to the effective groups; the sums over any thermal layers remain in the continuations. The construction thus corrects the interpretation of the product in the CCM~\cite{ConstructiveCavity}: the law of the input processes is a product, and ancestor assignments reconstruct the physical measure of the RPC with ultrametric spin overlaps. The input intensities are governed by the ratios of the exponents; the exponents of the recursive averages remain those of the condensed groups.

\section{Cavity formula and constructive variables}
\label{sec:cavity-rost}

We connect the hierarchical representation of the reservoir to the Aizenman--Sims--Starr formula to physically justify the identification of the constructive cavity variables with the ROSt~\cite{AizenmanSimsStarr,SimplifiedParisi}. We derive the two fresh Gaussian fields associated with the addition of a spin and the correction of the reservoir normalization, then transport the fields and the measure together to the ground state coordinates. The decomposition into layers and blankets allows their covariances to be calculated from the physical overlaps. We separate the components common to the fiber from the fluctuations due to the defects and specify their contribution to the cavity formula~\cite{RSBWithoutReplicas}. The geometry shared by the energies and the fields motivates the joint identification of the masses and the Gaussian variables with the ROSt. We then evaluate the mean logarithms of the two contributions on the cascade, obtaining the Parisi recursion and the quadratic correction; minimization of the functional and refinement of the levels complete the calculation of the SK pressure. Finally, we derive the tilt between the Blanket measure and the Gibbs measure and show how, by fixing the refined profiles of the stepwise model, it becomes constant on the fiber and allows the two conditional measures to be substituted for each other in the cavity formula~\cite{RSBWithoutReplicas,BlanketTheory}.

\subsection{Cavity formula in the form of the constructive method}
\label{subsec:formula-di-cavita-nella-forma-del-metodo-costruttivo}

We consider the SK reservoir with Hamiltonian~\eqref{eq:hamiltonian} and Gibbs measure~\eqref{eq:gibbs-measure}. We introduce two families of standard Gaussians, independent of each other and of the reservoir disorder. The tilde denotes this fresh noise. With the normalization of Ref.~\cite[Eqs.~(35)--(36)]{SimplifiedParisi}, we define the linear field
\begin{equation}
 \tilde x(\sigma_V)=\sqrt{\frac2N}\sum_{i\in V}\tilde J_i\sigma_i
 \label{eq:fresh-cavity-linear-field}
\end{equation}
and the correction field
\begin{equation}
 \tilde y(\sigma_V)=\frac1N\sum_{(i,j)\in W}\sigma_i\tilde J_{ij}\sigma_j.
 \label{eq:fresh-cavity-quadratic-field}
\end{equation}
The factor of two in the first field accounts for the two orientations of the bonds incident on the new spin. The second field uses all ordered pairs, as does the Hamiltonian. For a fixed configuration, their variances are two and one, respectively.

The cavity functional is
\begin{equation}
 A(\mu)=\log2+\log\left\langle\cosh[\beta\tilde x(\sigma_V)]\right\rangle_\mu-\log\left\langle e^{\beta\tilde y(\sigma_V)}\right\rangle_\mu.
 \label{eq:cavity-functional}
\end{equation}
This is the form of Ref.~\cite[Eq.~(46)]{SimplifiedParisi}, with the amplitudes appropriate to~\eqref{eq:hamiltonian}. Both contributions use the same reservoir measure. The hyperbolic cosine comes from the sum over the two values of the new spin; the exponential corrects the change in normalization of the preexisting interactions.

The calculation at finite volume specifies the rescaling. Separating the new spin and coupling the fresh noise, we have in distribution
\begin{equation}
 H(\sigma_{V\cup\{N+1\}})=\sqrt{\frac N{N+1}}\left[H(\sigma_V)+\sigma_{N+1}\tilde x(\sigma_V)\right]+\frac{\tilde J_{N+1,N+1}}{\sqrt{N+1}}.
 \label{eq:cavity-add-one-spin}
\end{equation}
The Gaussian field on the configurations
\begin{equation}
 \sqrt{\frac N{N+1}}\left[H(\sigma_V)-\tilde y(\sigma_V)\right]\stackrel d=H(\sigma_V)
 \label{eq:cavity-reservoir-coupling}
\end{equation}
instead has the same law as the original reservoir: the sum of the coefficient variances exactly restores the normalization of the system with $N$ spins. The sign chosen for the correction field gives the positive exponential in~\eqref{eq:cavity-functional}.

Dividing the two partition functions by that of the common reservoir, the inverse temperature appearing in the measure and in both contributions is
\begin{equation}
 \beta\sqrt{\frac N{N+1}}\longrightarrow\beta.
 \label{eq:cavity-finite-size-temperature}
\end{equation}
The new diagonal term is independent of the spins and has zero mean. Equations~\eqref{eq:cavity-add-one-spin} and~\eqref{eq:cavity-reservoir-coupling} therefore identify the difference between the means of the logarithms of the partition functions with the mean of the rescaled functional. Summing the increments and dividing by the volume recovers the pressure per spin, whose thermodynamic limit exists~\cite{GuerraToninelliLimit}:
\begin{equation}
 p=\lim_{N\to\infty}\frac1N\mathbb E\log Z.
 \label{eq:thermodynamic-pressure}
\end{equation}
Recomposition through the increments is the principle of Refs.~\cite{AizenmanSimsStarrOriginal} and~\cite[Sec.~2.1]{AizenmanSimsStarr}. Equation~\eqref{eq:cavity-functional} retains the random functional before averaging over the noise. The decomposition in Section~\ref{sec:blanket-framework} connects this pressure to that of the interface-only model in the subsequent refinement.

\subsection{Transport of the cavity variables to the rectified coordinates}
\label{subsec:trasporto-delle-variabili-di-cavita-nelle-coordinate-rettificate}

We simultaneously transport the reservoir measure and the two fields through the rectification~\eqref{eq:ground-state-spin-map}. With the reservoir disorder fixed, the reference is also fixed. The measure in the new coordinates is
\begin{equation}
 \mu^*(\sigma_V^*)=\mu(\sigma_V^*\circ\omega_V).
 \label{eq:rectified-cavity-reservoir-measure}
\end{equation}
The same identities hold for the interface-only measure~\eqref{eq:rectified-gibbs-measure}. We absorb the reference signs into the fresh coefficients:
\begin{equation}
 \tilde J_i^*=\omega_i\tilde J_i.
 \label{eq:rectified-linear-fresh-noise}
\end{equation}
\begin{equation}
 \tilde J_{ij}^*=\omega_i\tilde J_{ij}\omega_j.
 \label{eq:rectified-quadratic-fresh-noise}
\end{equation}
The symmetry of the Gaussians preserves their joint law: the two rectified families remain independent standard Gaussians, and this conditional law is the same for every realization of the reservoir. The rectified fresh noise is therefore also independent of the reference and of the energetic weights.

The transported fields are
\begin{equation}
 \tilde x^*(\sigma_V^*)=\tilde x(\sigma_V^*\circ\omega_V)=\sqrt{\frac2N}\sum_{i\in V}\tilde J_i^*\sigma_i^*.
 \label{eq:rectified-cavity-linear-field}
\end{equation}
\begin{equation}
 \tilde y^*(\sigma_V^*)=\tilde y(\sigma_V^*\circ\omega_V)=\frac1N\sum_{(i,j)\in W}\sigma_i^*\tilde J_{ij}^*\sigma_j^*.
 \label{eq:rectified-cavity-quadratic-field}
\end{equation}
The bijectivity of the change of coordinates preserves each sum entering the functional:
\begin{equation}
 \left\langle\cosh[\beta\tilde x(\sigma_V)]\right\rangle_\mu=\left\langle\cosh[\beta\tilde x^*(\sigma_V^*)]\right\rangle_{\mu^*}.
 \label{eq:rectified-linear-insertion}
\end{equation}
\begin{equation}
 \left\langle e^{\beta\tilde y(\sigma_V)}\right\rangle_\mu=\left\langle e^{\beta\tilde y^*(\sigma_V^*)}\right\rangle_{\mu^*}.
 \label{eq:rectified-quadratic-insertion}
\end{equation}
The two identities hold together, on the same states and with the same weights. The cavity calculation can therefore be developed on the rectified kernel while exactly preserving the functional~\eqref{eq:cavity-functional}. The configurations and the defects determine the geometry common to the reservoir energies and the constructive fields; the coefficients of the constructive fields come from the two fresh families just defined.

\subsection{Constructive cavity variables}
\label{subsec:variabili-di-cavita-costruttive}

The partition of the sites and bonds decomposes the two fresh fields into constructive contributions, following Ref.~\cite[Eqs.~(49), (51)--(53)]{SimplifiedParisi}. In the rectified coordinates we define
\begin{equation}
 \tilde z_\ell^*(\sigma_{V_\ell}^*)=\frac1{\sqrt{N_\ell}}\sum_{i\in V_\ell}\tilde J_i^*\sigma_i^*.
 \label{eq:constructive-linear-block}
\end{equation}
\begin{equation}
 \tilde g_\ell^*(\sigma_{Q_\ell}^*)=\frac1{\sqrt{|W_\ell|}}\sum_{(i,j)\in W_\ell}\sigma_i^*\tilde J_{ij}^*\sigma_j^*.
 \label{eq:constructive-quadratic-block}
\end{equation}
The field decompositions are exact:
\begin{equation}
 \tilde x^*(\sigma_V^*)=\sum_{\ell=1}^L\sqrt{2p_\ell}\,\tilde z_\ell^*(\sigma_{V_\ell}^*).
 \label{eq:constructive-linear-sum}
\end{equation}
\begin{equation}
 \tilde y^*(\sigma_V^*)=\sum_{\ell=1}^L\sqrt{q_\ell^2-q_{\ell-1}^2}\,\tilde g_\ell^*(\sigma_{Q_\ell}^*).
 \label{eq:constructive-quadratic-sum}
\end{equation}
The amplitudes follow from the volumes of the fresh noise blocks. Each normalized contribution has unit variance. Distinct blocks use disjoint coefficients and are independent; the two linear and quadratic families are themselves independent. The correction field includes all blankets, including the cores, because it compensates for the normalization of the entire reservoir.

The calculation of the covariances follows directly from the Gaussian sums. Using the notation of Ref.~\cite[Def.~2, Eq.~(2.11)]{RSBWithoutReplicas}, we denote the overlap of the configurations by
\begin{equation}
 q(\sigma_V^*,\tau_V^*)=\frac1N\sum_{i\in V}\sigma_i^*\tau_i^*.
 \label{eq:physical-overlap-function}
\end{equation}
The following averages concern only the fresh noise, with the configurations and the reservoir disorder fixed. For the linear layer field we obtain
\begin{equation}
 \mathbb E\!\left[\tilde z_\ell^*(\sigma_{V_\ell}^*)\tilde z_\ell^*(\tau_{V_\ell}^*)\right]=\frac1{N_\ell}\sum_{i\in V_\ell}\sigma_i^*\tau_i^*.
 \label{eq:constructive-linear-block-covariance}
\end{equation}
For the blanket field,
\begin{equation}
 \mathbb E\!\left[\tilde g_\ell^*(\sigma_{Q_\ell}^*)\tilde g_\ell^*(\tau_{Q_\ell}^*)\right]=\frac{\left(\sum_{i\in Q_\ell}\sigma_i^*\tau_i^*\right)^2-\left(\sum_{i\in Q_{\ell-1}}\sigma_i^*\tau_i^*\right)^2}{|W_\ell|}.
 \label{eq:constructive-quadratic-block-covariance}
\end{equation}
The sum over the blankets recomposes the square of the total overlap. Equations~\eqref{eq:constructive-linear-sum} and~\eqref{eq:constructive-quadratic-sum} therefore give
\begin{equation}
 \mathbb E\!\left[\tilde x^*(\sigma_V^*)\tilde x^*(\tau_V^*)\right]=2q(\sigma_V^*,\tau_V^*).
 \label{eq:full-linear-cavity-covariance}
\end{equation}
\begin{equation}
 \mathbb E\!\left[\tilde y^*(\sigma_V^*)\tilde y^*(\tau_V^*)\right]=q(\sigma_V^*,\tau_V^*)^2.
 \label{eq:full-quadratic-cavity-covariance}
\end{equation}
These are the Gaussian covariances of the ROSt in the convention adopted~\cite[Sec.~2.3--2.4]{AizenmanSimsStarr}. The rectification preserves them by~\eqref{eq:rectified-overlap-invariance}.

In the construction of Section~\ref{sec:grem-universality}, two states with their last common ancestor at level $\ell$ coincide on the first layers, as in Figure~\ref{fig:kernel-grem-gerarchia}. In the subsequent ones, the mean intersection of two uniform supports gives the square of the magnetization. We denote the hierarchical values of the physical overlap by
\begin{equation}
 \rho_\ell:=q_\ell+\sum_{t>\ell}p_t m_t^2,\qquad \ell=0,\ldots,L.
 \label{eq:hierarchical-physical-overlap}
\end{equation}
Equations~\eqref{eq:extensive-tree-overlap} and~\eqref{eq:extensive-overlap-correction} uniformly control the difference from this value:
\begin{equation}
 \left|q(\sigma_V^{*\alpha},\sigma_V^{*\beta})-\rho_\ell\right|\leq4\sum_{t>\ell}p_t\epsilon_t^{3/2}.
 \label{eq:hierarchical-overlap-error}
\end{equation}
The initial shared contribution is the sum of the squared magnetizations weighted by the volumes. The connection between overlaps, volumes, and magnetizations has a precedent in the description of local magnetizations in terms of cells and ancestors in Ref.~\cite{MicrostructureUltrametricity}. Each new level adds to the overlap
\begin{equation}
 \rho_\ell-\rho_{\ell-1}=p_\ell(1-m_\ell^2).
 \label{eq:hierarchical-overlap-increment}
\end{equation}
The covariance of the linear field gains twice this increment. That of the correction field gains the difference between the squares of the cumulative values:
\begin{equation}
 \rho_\ell^2-\rho_{\ell-1}^2.
 \label{eq:hierarchical-quadratic-increment}
\end{equation}
These quantities geometrically determine the amplitudes of the ROSt fields. The complete energy laws in~\eqref{eq:gaussian-grem-complete-energy}, with their amplitudes and the populations, instead determine the mass exponents through the cost~\eqref{eq:grem-complete-energy-rate}. The kernel brings the two sets of data together: the configurations fix the overlaps and the weights fix the widths of the columns.

\subsection{Common component of the cavity field}
\label{subsec:componente-comune-del-campo-di-cavita}

On the fiber with fixed magnetizations, the fresh field contains a common component and a part that varies with the defects. This separation specifies the role of the fluctuations considered in Ref.~\cite[Lemma~13]{RSBWithoutReplicas} and how they enter the complete cavity formula. We keep the energetic disorder and the reservoir weights fixed during the calculation over the fresh noise.

The uniform mean of the linear field on the fiber is
\begin{equation}
 \left\langle\tilde x^*(\sigma_V^*)\right\rangle_{m_1,\ldots,m_L}=\sum_{\ell=1}^L\sqrt{2p_\ell}\,m_\ell\tilde z_\ell^*(1_{V_\ell}).
 \label{eq:common-linear-cavity-field}
\end{equation}
The centered field is exactly
\begin{equation}
 \tilde x^*(\sigma_V^*)-\left\langle\tilde x^*(\sigma_V^*)\right\rangle_{m_1,\ldots,m_L}=\sqrt{\frac2N}\sum_{\ell=1}^L\sum_{i\in V_\ell}\tilde J_i^*(\sigma_i^*-m_\ell).
 \label{eq:centered-linear-cavity-field}
\end{equation}
In each layer the sum of the centered spins is zero. The coefficient vector of the common component is therefore orthogonal to that of each centered field. Gaussianity implies the independence of the common component from the entire centered family. The variances with respect to the fresh noise are
\begin{equation}
 \mathbb E\!\left[\left\langle\tilde x^*(\sigma_V^*)\right\rangle_{m_1,\ldots,m_L}^{\,2}\right]=2\sum_{\ell=1}^L p_\ell m_\ell^2=2\rho_0.
 \label{eq:common-linear-cavity-variance}
\end{equation}
\begin{equation}
 \mathbb E\!\left[\left(\tilde x^*(\sigma_V^*)-\left\langle\tilde x^*(\sigma_V^*)\right\rangle_{m_1,\ldots,m_L}\right)^2\right]=2\sum_{\ell=1}^L p_\ell(1-m_\ell^2)=2(1-\rho_0).
 \label{eq:centered-linear-cavity-variance}
\end{equation}
Their sum recovers the complete field's variance of two. The first coincides with the initial covariance of the hierarchical description~\eqref{eq:hierarchical-physical-overlap}; the second is distributed among the increments~\eqref{eq:hierarchical-overlap-increment}.

The linear contribution conditioned on the fiber retains the complete Gibbs weights:
\begin{equation}
 \log\frac{\left\langle\bar\mu^*(\sigma_V^*)\,2\cosh[\beta\tilde x^*(\sigma_V^*)]\right\rangle_{m_1,\ldots,m_L}}{\left\langle\bar\mu^*(\sigma_V^*)\right\rangle_{m_1,\ldots,m_L}}.
 \label{eq:common-field-complete-insertion}
\end{equation}
The common component acts as a Gaussian field on the new spin. It remains fixed while the reservoir states are summed over and the fluctuations of the subsequent levels are integrated. In the evaluation of the cascade average, it is integrated last, after the recursions of the condensed levels~\cite[Eqs.~(10), (14)]{MezardParisiVirasoroCavity}. In this way the hyperbolic cosine receives the complete field of~\eqref{eq:rectified-cavity-linear-field}.

The correction field also has a common mean on the fiber. The symmetry of the sampling makes the scalar product between its coefficient vector and the mean vector constant on the fiber; the residual is therefore orthogonal to the mean. The mean and the residual family are independent Gaussians. From the identities~\eqref{eq:shell-spin-mean}--\eqref{eq:shell-spin-pair}, with the number of layers and the volume fractions fixed, it follows that
\begin{equation}
 \mathbb E\!\left[\left\langle\tilde y^*(\sigma_V^*)\right\rangle_{m_1,\ldots,m_L}^{\,2}\right]\longrightarrow\left(\sum_{\ell=1}^L p_\ell m_\ell^2\right)^2=\rho_0^2.
 \label{eq:common-quadratic-cavity-variance}
\end{equation}
In the exponential contribution the common component separates exactly:
\begin{equation}
 \begin{aligned}
 &\log\frac{\left\langle\bar\mu^*(\sigma_V^*)e^{\beta\tilde y^*(\sigma_V^*)}\right\rangle_{m_1,\ldots,m_L}}{\left\langle\bar\mu^*(\sigma_V^*)\right\rangle_{m_1,\ldots,m_L}}=\beta\left\langle\tilde y^*(\sigma_V^*)\right\rangle_{m_1,\ldots,m_L}\\
 &\quad+\log\frac{\left\langle\bar\mu^*(\sigma_V^*)\exp\!\left[\beta\left(\tilde y^*(\sigma_V^*)-\left\langle\tilde y^*(\sigma_V^*)\right\rangle_{m_1,\ldots,m_L}\right)\right]\right\rangle_{m_1,\ldots,m_L}}{\left\langle\bar\mu^*(\sigma_V^*)\right\rangle_{m_1,\ldots,m_L}}.
 \end{aligned}
 \label{eq:common-quadratic-insertion}
\end{equation}
The first term is present in the random functional and has zero mean over the fresh noise. The remaining part follows the quadratic increments~\eqref{eq:hierarchical-quadratic-increment}.

In the nearly aligned limit, the variance of the centered field~\eqref{eq:centered-linear-cavity-variance} decreases, while the common variance tends to two. When the fluctuations vanish, the linear contribution becomes the logarithm of twice the hyperbolic cosine of the common field alone. For nonzero fluctuations, its response remains described by~\eqref{eq:common-field-complete-insertion}. At zero external field, global inversion of the reservoir changes the sign of the linear field and preserves the quadratic field; both contributions are therefore equal in the two symmetric sectors and their weights add.

The REM rescaling in Subsection~\ref{subsec:entropie-pendenze-e-livelli-effettivi} accounts for the population of states and the amplitude of the energies. The separation~\eqref{eq:common-linear-cavity-field}--\eqref{eq:centered-linear-cavity-field} instead keeps track of the components of the field entering the functional. In the functional both carry the physical temperature of~\eqref{eq:cavity-functional}; the common component constitutes the first Gaussian datum of the ROSt and the fluctuations build the subsequent increments.

\subsection{Convergence to the ROSt}
\label{subsec:ansatz-di-convergenza-alla-rost}

We connect the GREM representation of Section~\ref{sec:grem-universality} to the constructive fields through the joint law of the masses and the overlaps. The geometry of the defects determines the correlations of the fresh fields; Gaussianity then allows their distribution to be identified from the covariances already calculated. This is the connection to the ROSt of Ref.~\cite[Sec.~2.3--2.4, 5.3]{AizenmanSimsStarr} required by the constructive formulation of Ref.~\cite{SimplifiedParisi}.

In the representation of Section~\ref{sec:grem-universality}, the masses follow from the complete branch energies and the overlaps from their shared defects. The energy increments inherited by the descendants produce the prefix masses; the same configurations determine the value~\eqref{eq:hierarchical-physical-overlap}. The derivation of the GREM representative uses the common linear increments~\eqref{eq:dilute-common-law-increment} on the same hierarchical configurations, with the bilinear contribution controlled in the dilute free-energy comparison~\eqref{eq:dilute-bilinear-free-energy-bound}. Its physical motivation is the mechanism of Subsection~\ref{subsec:rappresentazione-grem}: states with the same prefix reuse the same defects and the same energy contributions, while distinct nodes add new arrangements. Subsection~\ref{subsec:passaggio-al-regime-estensivo} extends the sampling to exponential populations and specifies the approximation of the local energy laws by data common to the level. The selection of the extreme energies finally determines the complete masses of Subsections~\ref{subsec:cascata-grem-e-prodotto-di-ppp-a-due-livelli} and~\ref{subsec:prodotto-di-ppp-ed-equivalenza-in-distribuzione-con-la-cascata}.

We first fix a physical fiber and the parameters of its representation, as in Subsection~\ref{subsec:entropie-pendenze-e-livelli-effettivi}. For the normalized cascade masses we use the GREM notation:
\begin{equation}
 \sum_\alpha\bar\mu_{\mathrm{GREM}}^*(\sigma_V^\alpha)=1.
 \label{eq:rost-complete-mass-normalization}
\end{equation}
Each mass includes all continuations of its branch. The state indices follow~\eqref{eq:state-path-indices}; the merged layers share the index of the effective group. Any thermal suffix is summed over within the continuations.

Conditionally on the kernel, the constructive fields are linear combinations of the fresh noise. Their joint law is therefore Gaussian already at finite volume and is determined by~\eqref{eq:full-linear-cavity-covariance} and~\eqref{eq:full-quadratic-cavity-covariance}. The two families have zero mutual covariance because they use independent noise. For any finite number of states, the convergence of their masses and their overlaps therefore also determines the joint convergence of the masses and the two fields. The energetic selection of the configurations changes the weights and the overlaps; with the kernel fixed, the law of the fresh noise remains that of Subsection~\ref{subsec:trasporto-delle-variabili-di-cavita-nelle-coordinate-rettificate}.

We explicitly construct the fields of the representative. As in Refs.~\cite[Eqs.~(52)--(55)]{SimplifiedParisi} and~\cite[Eqs.~(5.29)--(5.30)]{AizenmanSimsStarr}, we associate two families of standard Gaussians with the nodes:
\begin{equation}
 \tilde z_\ell^{\alpha_1\cdots\alpha_\ell}\sim\mathcal N(0,1).
 \label{eq:rost-normalized-linear-increments}
\end{equation}
\begin{equation}
 \tilde g_\ell^{\alpha_1\cdots\alpha_\ell}\sim\mathcal N(0,1).
 \label{eq:rost-normalized-quadratic-increments}
\end{equation}
The Gaussians of distinct nodes are independent; the descendants of the same node reuse its increment. The two families are independent of each other, of the masses, and of the common components. For the latter we retain the notation of Subsection~\ref{subsec:componente-comune-del-campo-di-cavita}, with the limiting Gaussian laws and the variances of~\eqref{eq:common-linear-cavity-variance} and~\eqref{eq:common-quadratic-cavity-variance}.

The complete linear field is
\begin{equation}
 \tilde x(\sigma_V^\alpha)=\left\langle\tilde x^*(\sigma_V^*)\right\rangle_{m_1,\ldots,m_L}+\sum_{\ell=1}^L\sqrt{2(\rho_\ell-\rho_{\ell-1})}\,\tilde z_\ell^{\alpha_1\cdots\alpha_\ell}.
 \label{eq:rost-complete-linear-field}
\end{equation}
For the correction field we obtain
\begin{equation}
 \begin{aligned}
 \tilde y(\sigma_V^\alpha)&=\left\langle\tilde y^*(\sigma_V^*)\right\rangle_{m_1,\ldots,m_L}\\
 &\quad+\sum_{\ell=1}^L\sqrt{\rho_\ell^2-\rho_{\ell-1}^2}\,\tilde g_\ell^{\alpha_1\cdots\alpha_\ell}.
 \end{aligned}
 \label{eq:rost-complete-quadratic-field}
\end{equation}
The amplitudes are those of~\eqref{eq:hierarchical-overlap-increment} and~\eqref{eq:hierarchical-quadratic-increment}. The sum of the variances, including the common parts, gives two and one, respectively. For two states with their last common ancestor at level $\ell$, the covariance is obtained by summing only the shared increments. The telescoping sums give twice the value~\eqref{eq:hierarchical-physical-overlap} for the linear field and its square for the correction field. Equations~\eqref{eq:rost-complete-linear-field} and~\eqref{eq:rost-complete-quadratic-field} therefore realize the complete Gaussian law required by the kernel overlaps.

The renewal of the defects thus has a direct counterpart in the cavity fields: when two branches separate, the new Gaussian increments are distinct; the contributions of the ancestors and the initial component remain shared. The identification concerns the recomposed complete fields. The geometric coefficients come from the overlaps, while the complete energy laws and the populations of Subsection~\ref{subsec:entropie-pendenze-e-livelli-effettivi} determine the mass exponents. These data are brought together on the same states of the representative.

The matching of the two contributions, on the fiber considered, therefore takes the form
\begin{equation}
 \frac{\left\langle\bar\mu^*(\sigma_V^*)\cosh[\beta\tilde x^*(\sigma_V^*)]\right\rangle_{m_1,\ldots,m_L}}{\left\langle\bar\mu^*(\sigma_V^*)\right\rangle_{m_1,\ldots,m_L}}\overset d\simeq\sum_\alpha\bar\mu_{\mathrm{GREM}}^*(\sigma_V^\alpha)\cosh[\beta\tilde x(\sigma_V^\alpha)].
 \label{eq:rost-linear-insertion-identification}
\end{equation}
\begin{equation}
 \frac{\left\langle\bar\mu^*(\sigma_V^*)e^{\beta\tilde y^*(\sigma_V^*)}\right\rangle_{m_1,\ldots,m_L}}{\left\langle\bar\mu^*(\sigma_V^*)\right\rangle_{m_1,\ldots,m_L}}\overset d\simeq\sum_\alpha\bar\mu_{\mathrm{GREM}}^*(\sigma_V^\alpha)e^{\beta\tilde y(\sigma_V^\alpha)}.
 \label{eq:rost-quadratic-insertion-identification}
\end{equation}
The two relations express a single joint approximation: on the left they share the physical configurations and their weights, on the right the cascade and its ancestors. The approximation symbol refers to the GREM description of Section~\ref{sec:grem-universality}. Within this description, the passage from the overlaps to the Gaussian fields is determined by the covariances.

After the ordinary average over any thermal continuations, a finite number of complete masses of the condensed prefixes collects a fraction of the measure arbitrarily close to one. At fixed temperature, the Gaussian moments of the two contributions are uniform in the states and the volume, because the field variances are fixed. The mean contribution of an omitted set of states is therefore proportional to its total mass. One can thus pass from the connection for a finite number of states to the complete sums. For the logarithms, Jensen's inequality controls the negative part of the exponential contribution through the average of the field; the Gaussian moments control the positive part. The hyperbolic cosine is at least one. These controls also allow passage to the averages of the functional at fixed temperature, following the calculation of Ref.~\cite[Lemma~2.2]{AizenmanSimsStarr}. The subsequent low-temperature limit concerns the physical description of the defects and the parameters of the hierarchy.

The recomposition of the fibers follows~\eqref{eq:gibbs-fibre-weight} and~\eqref{eq:observable-disintegration}: each conditional contribution is multiplied by the complete weight of the fiber. In the presence of several contributing profiles, these weights and the same fresh noise are retained across the different fibers. The identification with the ROSt therefore concerns the masses and the overlaps of the physical kernel thus recomposed. The following calculation evaluates the hierarchical representative at fixed parameters; the choice of the parameters and the refinement of the levels complete the connection to the thermodynamic functional.

\subsection{Cavity contributions and Parisi recursion}
\label{subsec:contributi-di-cavita-e-ricorsione-di-parisi}

We calculate the functional and its average on the ROSt just constructed. The construction of Subsection~\ref{subsec:prodotto-di-ppp-ed-equivalenza-in-distribuzione-con-la-cascata} transports the masses, the ancestors, and the two Gaussian families together. For each terminal point, the indices of the preceding levels are the ancestors assigned only once; the descendants of the same node reuse its increment. Equation~\eqref{eq:independent-marked-average}, applied to the two contributions with the same assignment, gives
\begin{equation}
 A(\bar\mu_{\mathrm{GREM}}^*)\stackrel d=\log2+\log\frac{\displaystyle\sum_{\alpha_L}\xi_L^{\alpha_L}\cosh[\beta\tilde x(\sigma_V^{\alpha_1\cdots\alpha_L})]}{\displaystyle\sum_{\alpha_L}\xi_L^{\alpha_L}\exp[\beta\tilde y(\sigma_V^{\alpha_1\cdots\alpha_L})]}.
 \label{eq:marked-input-cavity-functional}
\end{equation}
The fields retain the common components and the amplitudes of~\eqref{eq:rost-complete-linear-field} and~\eqref{eq:rost-complete-quadratic-field}. The global normalization is the same in the two sums and cancels in the ratio. The identity preserves the random functional before averaging.

The calculation of the average uses the independent PPPs of~\eqref{eq:independent-local-rem-processes}, with the totals $Z_\ell$ defined in Subsection~\ref{subsec:cascata-grem-e-prodotto-di-ppp-a-due-livelli}. We consider a positive function $f$ of a standard Gaussian mark, with a moment of the required order and integrable logarithms. The functions of the two contributions satisfy these conditions. We keep any common fields and the preceding increments fixed. Multiplying each point of the input process by its mark gives the intensity
\begin{equation}
 \left\{Z_\ell\xi_\ell^{\alpha_\ell}f(\tilde z_\ell^{\alpha_\ell}):\alpha_\ell\geq1\right\},\qquad \lambda_\ell\mathbb E\!\left[f(\tilde z_\ell)^{\lambda_\ell}\right]x^{-1-\lambda_\ell}\,dx.
 \label{eq:marked-poisson-intensity}
\end{equation}
Here the marks are independent of each other and of the process, and the average in the coefficient concerns a single mark. The change of scale of the intensity implies
\begin{equation}
 \sum_{\alpha_\ell}Z_\ell\xi_\ell^{\alpha_\ell}f(\tilde z_\ell^{\alpha_\ell})\stackrel d=\left(\mathbb E\!\left[f(\tilde z_\ell)^{\lambda_\ell}\right]\right)^{1/\lambda_\ell}Z_\ell.
 \label{eq:marked-poisson-total-law}
\end{equation}
The formula specifies the law of the individual sum. Taking the logarithm and subtracting the mean logarithm of the unmarked total gives
\begin{equation}
 \mathbb E\log\left[\sum_{\alpha_\ell}\xi_\ell^{\alpha_\ell}f(\tilde z_\ell^{\alpha_\ell})\right]=\frac1{\lambda_\ell}\log\mathbb E\!\left[f(\tilde z_\ell)^{\lambda_\ell}\right].
 \label{eq:normalized-poisson-mean-log}
\end{equation}
The subtraction uses the linearity of the average: the total and the marked sum come from the same process. This is the identity that enters the recursive evaluation of Refs.~\cite[Sec.~5.3, 6.2]{AizenmanSimsStarr} and~\cite[Eqs.~(7)--(8)]{SimplifiedParisi}.

The two-level case shows how the input exponents recover those of the cascade. Conditionally on the first process and its marks, each point of the second process receives a parent with the probabilities~\eqref{eq:two-level-parent-assignment} and a new independent Gaussian mark. For a positive function of the two marks, equation~\eqref{eq:normalized-poisson-mean-log} at the second level gives
\begin{equation}
 \begin{aligned}
 &\mathbb E\log\left[\sum_{\alpha_2}\xi_2^{\alpha_2}f(\tilde z_1^{\alpha_1},\tilde z_2^{\alpha_2})\right]\\
 &\quad=\frac1{\lambda_2}\mathbb E\log\frac{\displaystyle\sum_{\alpha_1}(\xi_1^{\alpha_1})^{\lambda_2}\,\mathbb E_2\!\left[f(\tilde z_1^{\alpha_1},\tilde z_2)^{\lambda_2}\right]}{\displaystyle\sum_{\beta_1}(\xi_1^{\beta_1})^{\lambda_2}}.
 \end{aligned}
 \label{eq:two-level-conditional-insertion}
\end{equation}
On the left the first index is the parent of the point being summed over. On the right the outer average concerns the first process and its marks; $\mathbb E_2$ integrates only the second mark. The points of the first process raised to the power $\lambda_2$ have exponent $\lambda_1/\lambda_2$, according to~\eqref{eq:powered-input-processes}. A second application of~\eqref{eq:normalized-poisson-mean-log} gives
\begin{equation}
 \mathbb E\log\left[\sum_{\alpha_2}\xi_2^{\alpha_2}f(\tilde z_1^{\alpha_1},\tilde z_2^{\alpha_2})\right]=\frac1{\lambda_1}\log\mathbb E_1\!\left[\left(\mathbb E_2\!\left[f(\tilde z_1,\tilde z_2)^{\lambda_2}\right]\right)^{\lambda_1/\lambda_2}\right].
 \label{eq:two-level-parisi-average}
\end{equation}
The same operation is repeated from the last level toward the first. At each step the assignments and the noise of the new nodes are independent conditionally on the preceding levels. The exponent ratios govern the input probabilities; the composition of the averages recovers the exponents $\lambda_\ell$.

For the linear contribution we use the recursive variables $Y_\ell$ and the Gaussian average $\mathbb E_\ell$ of Ref.~\cite[Eqs.~(7)--(8)]{SimplifiedParisi}. The common component remains fixed. On a branch, the terminal condition is
\begin{equation}
 Y_L=\cosh\!\left\{\beta\left[\left\langle\tilde x^*(\sigma_V^*)\right\rangle_{m_1,\ldots,m_L}+\sum_{\ell=1}^L\sqrt{2(\rho_\ell-\rho_{\ell-1})}\,\tilde z_\ell\right]\right\}.
 \label{eq:parisi-terminal-common-field}
\end{equation}
The Gaussians $\tilde z_\ell$ are the independent increments along the branch; $\mathbb E_\ell$ integrates only $\tilde z_\ell$. The preceding increments and the common component remain fixed in the recursion~\cite[Eqs.~(9), (13)]{MezardParisiVirasoroCavity}
\begin{equation}
 Y_{\ell-1}=\left(\mathbb E_\ell Y_\ell^{\lambda_\ell}\right)^{1/\lambda_\ell},\qquad \ell=L,\ldots,1.
 \label{eq:parisi-power-recursion}
\end{equation}
After the last step, $Y_0$ depends only on the common field, whose variance is given by~\eqref{eq:common-linear-cavity-variance}. Integrating this field last gives
\begin{equation}
 \mathbb E\log\left[\sum_\alpha\bar\mu_{\mathrm{GREM}}^*(\sigma_V^\alpha)\cosh[\beta\tilde x(\sigma_V^\alpha)]\right]=\mathbb E\log Y_0.
 \label{eq:linear-insertion-mean-recursion}
\end{equation}
The average on the left includes the masses and the fresh noise; the one on the right concerns only the common field remaining after the recursion. Its contribution is equivalent to the initial step with exponent zero of the Parisi functional.

The quadratic contribution is calculated with the same recursive averages. The common component separates from the logarithm as in~\eqref{eq:common-quadratic-insertion} and has zero mean. For each increment of~\eqref{eq:rost-complete-quadratic-field}, the Gaussian moment gives
\begin{equation}
 \frac1{\lambda_\ell}\log\mathbb E\exp\!\left\{\beta\lambda_\ell\sqrt{\rho_\ell^2-\rho_{\ell-1}^2}\,\tilde g_\ell\right\}=\frac{\beta^2\lambda_\ell}{2}\left[\rho_\ell^2-\rho_{\ell-1}^2\right].
 \label{eq:quadratic-increment-gaussian-average}
\end{equation}
The exponential transforms the sum of the increments into a product; each step therefore adds the contribution just calculated. It follows that
\begin{equation}
 \mathbb E\log\left[\sum_\alpha\bar\mu_{\mathrm{GREM}}^*(\sigma_V^\alpha)e^{\beta\tilde y(\sigma_V^\alpha)}\right]=\frac{\beta^2}{2}\sum_{\ell=1}^L\lambda_\ell\left[\rho_\ell^2-\rho_{\ell-1}^2\right].
 \label{eq:quadratic-insertion-mean-recursion}
\end{equation}

Combining the two contributions, the average of the functional is
\begin{equation}
 \mathbb E A(\bar\mu_{\mathrm{GREM}}^*)=\log2+\mathbb E\log Y_0-\frac{\beta^2}{2}\sum_{\ell=1}^L\lambda_\ell\left[\rho_\ell^2-\rho_{\ell-1}^2\right].
 \label{eq:physical-overlap-parisi-functional}
\end{equation}
\begingroup\emergencystretch=1em This is the Parisi functional~\cite{ParisiOrderParameter} in the form of Ref.~\cite[Eqs.~(7)--(8)]{SimplifiedParisi}, written in terms of the physical overlaps~\eqref{eq:hierarchical-physical-overlap}. The linear covariance is twice the overlap and the quadratic one is its square; the coefficient of the correction follows from these conventions. The equivalence in distribution~\eqref{eq:marked-input-cavity-functional} preserves the random functional, while~\eqref{eq:physical-overlap-parisi-functional} calculates its average. This is the averaged result that replaces the step in Eqs.~(35)--(40) of Ref.~\cite{ConstructiveCavity}, discussed at the beginning of Subsection~\ref{subsec:cascata-grem-e-prodotto-di-ppp-a-due-livelli}.\par\endgroup

For the merged layers, the group increments are recomposed and its exponent is used. Equation~\eqref{eq:parisi-power-recursion} can also be applied to the original layers by assigning them the same exponent: consecutive averages with the same exponent compose into the average over the entire group. In any thermal suffix, one sets $\lambda_\ell=1$, and the recursion becomes the ordinary Gaussian average. This average is performed before the condensed steps and keeps the continuations within each contribution. The extreme values of the exponents are obtained by continuity; when the exponent tends to zero, the logarithm of the power mean tends to the mean logarithm.

The selection of the parameters for SK now takes place on the functional~\eqref{eq:physical-overlap-parisi-functional}, according to the variational principle of Refs.~\cite[Eq.~(6)]{SimplifiedParisi} and~\cite[Sec.~6.1]{AizenmanSimsStarr}. The variable data are the ordered values of the physical overlaps and the mass exponents. The volume fractions and the magnetizations provide their geometric representation of Section~\ref{sec:grem-universality}; the populations and the energy laws determine the exponents through the balance of Subsection~\ref{subsec:entropie-pendenze-e-livelli-effettivi}. For a finite hierarchy, an approximate variational value is obtained. The minimum over the parameters of the hierarchical representation and the subsequent refinement of the levels recover the SK pressure in the convention adopted~\cite{GuerraRSBBound,TalagrandParisiFormula}. The cascade geometry and its exponents are thus selected after the derivation of the functional.

\subsection{Operational form of the CCM}
\label{subsec:operational-form-of-the-ccm}

The preceding subsection derives the Parisi functional from the physical overlaps and the masses of the hierarchical states. The operational CCM of Refs.~\cite{SimplifiedParisi,RSBWithoutReplicas} calculates the coefficients of the cavity fields from the dimensions of the layers and blankets. We now explain the physical correspondence between these descriptions and show how the construction based on block dimensions reproduces the same functional.

The reservoir enters the cavity formula through its state masses and the correlations between the fields generated by its configurations. A newly added spin couples to these configurations through fresh Gaussian coefficients. Their overlaps determine the covariance of the linear fields, while the squared overlaps determine the covariance of the correction fields. Since the two field families are Gaussian and independent of each other, their joint law is determined by these covariances. Preserving the joint distribution of the masses and the overlap matrix therefore preserves the distribution of the random cavity functional~\cite[Secs.~2.3--2.4]{AizenmanSimsStarr}.

Within the hierarchical representation developed above, this correspondence can be understood directly from the physical fiber. Two states with the same prefix have identical spins in the layers belonging to that prefix. In the subsequent layers, their alignment with the reference configuration contributes through the squared magnetizations. Equations~\eqref{eq:hierarchical-physical-overlap} and~\eqref{eq:hierarchical-overlap-increment} give two equivalent expressions for their overlap:
\begin{equation}
 \rho_\ell=q_\ell+\sum_{t>\ell}p_t m_t^2=\rho_0+\sum_{t\leq \ell}p_t(1-m_t^2).
 \label{eq:operational-overlap-recomposition}
\end{equation}
The first expression resolves the correlations over the physical layers: the shared prefix contributes its full volume, and the remaining layers contribute their mean alignment. The second collects the mean alignment of all layers into the common component and adds the fluctuations shared along the prefix. Each new common ancestor contributes the variance of the spin fluctuations in its layer, weighted by the layer volume.

The operational representation assigns these contributions to block dimensions. Choose a constructive partition with successive volume fractions
\begin{equation}
 \rho_0,\quad \rho_1-\rho_0,\quad\ldots,\quad\rho_L-\rho_{L-1}.
 \label{eq:operational-partition-fractions}
\end{equation}
These fractions are nonnegative and sum to one, so they can be realized by site blocks in the thermodynamic limit. Intervals of zero width are omitted. The common block, when present, represents the component shared by all states; the subsequent blocks represent the fluctuations inherited from the successive ancestors. The original state masses and ancestry are retained, and each interval between successive physical overlaps carries the exponent of the corresponding level. The common block is treated by the Gaussian average at exponent zero described in Subsection~\ref{subsec:contributi-di-cavita-e-ricorsione-di-parisi}.

For the calculation on this chosen partition, we reindex the intervals and use the existing notation: $L$ denotes their total number, $q_\ell$ their cumulative volume fractions, and $\lambda_\ell$ their averaging exponents. Thus the fractions used below describe the operational partition constructed from~\eqref{eq:operational-partition-fractions}. Its blocks represent the contributions to the correlations previously distributed over the physical layers.

Assign independent standard Gaussian increments to distinct nodes, with the coefficients calculated from the block volumes as in Eqs.~\eqref{eq:constructive-linear-sum}--\eqref{eq:constructive-quadratic-sum}. Descendants share the increments of their ancestors, and the Gaussian variable of the common block, when present, is shared by all states. The linear and quadratic noise families are independent. For two states whose last shared level is $\ell$, summing the variances of their shared increments gives
\begin{equation}
 \mathbb E\!\left[\tilde x(\sigma_V^\alpha)\tilde x(\sigma_V^\beta)\right]=2q_\ell.
 \label{eq:operational-linear-covariance}
\end{equation}
The corresponding blanket volumes give
\begin{equation}
 \mathbb E\!\left[\tilde y(\sigma_V^\alpha)\tilde y(\sigma_V^\beta)\right]=q_\ell^2.
 \label{eq:operational-quadratic-covariance}
\end{equation}
These averages concern the fresh noise at fixed hierarchy. By construction, the covariances coincide with those of the physical representative. The construction also reproduces the common-field variances in Eqs.~\eqref{eq:common-linear-cavity-variance} and~\eqref{eq:common-quadratic-cavity-variance}. The product of PPP constructed in Subsection~\ref{subsec:prodotto-di-ppp-ed-equivalenza-in-distribuzione-con-la-cascata} provides an equivalent realization of the masses and ancestry through global normalization and the triangular parent assignments in Eq.~\eqref{eq:multilevel-parent-assignment}. Each node inherits its parent's complete prefix, and all its descendants share the field increments associated with that prefix. The assignments determine which increments are shared, while the block dimensions in Eq.~\eqref{eq:operational-partition-fractions} determine their variances. Transporting the masses and both complete cavity fields through the same assignments therefore gives the same joint law as the RPC carrying the physical overlaps and preserves the random cavity functional in distribution. The two descriptions represent the same statistical response of the reservoir in the cavity formula.

The average functional can therefore be calculated using the block dimensions. We denote the recursive variables associated with the operational partition by $Y_\ell$. Along a branch, the terminal condition is
\begin{equation}
 Y_L=\cosh\!\left[\beta\sum_{\ell=1}^{L}\sqrt{2(q_\ell-q_{\ell-1})}\,\tilde z_\ell\right].
 \label{eq:operational-parisi-terminal}
\end{equation}
The Gaussian variables are the independent increments along the branch, including the common increment. For positive exponents, the recursion is~\eqref{eq:parisi-power-recursion}, with $\mathbb E_\ell$ integrating the corresponding Gaussian variable. Its extension to exponent zero is
\begin{equation}
 \log Y_{\ell-1}=\mathbb E_\ell\log Y_\ell,\qquad \lambda_\ell=0.
 \label{eq:operational-zero-exponent-step}
\end{equation}
The common linear field remains inside the hyperbolic cosine during the averages over the subsequent levels and is integrated last. The common correction field has zero mean, as established in Eq.~\eqref{eq:common-quadratic-insertion}. After all these averages, $Y_0$ is deterministic, and the functional becomes
\begin{equation}
 \mathbb E A(\bar\mu_{\mathrm{GREM}}^*)=\log2+\log Y_0-\frac{\beta^2}{2}\sum_{\ell=1}^{L}\lambda_\ell(q_\ell^2-q_{\ell-1}^2).
 \label{eq:operational-parisi-functional}
\end{equation}
Here $\log Y_0$ includes the final average over the common field and equals the contribution written as $\mathbb E\log Y_0$ in Eq.~\eqref{eq:physical-overlap-parisi-functional}. Equation~\eqref{eq:operational-parisi-functional} therefore expresses the same functional through the dimensions of the constructive blocks, with the physical temperature unchanged. The identification of the hierarchical ROSt functional with the Parisi functional is consistent with Ref.~\cite[Sec.~6.2]{AizenmanSimsStarr}.

The correspondence also explains how to use a finer geometric partition. Subdivide an interval and assign its exponent to every resulting subinterval. For two adjacent steps with the same positive exponent, applying the recursion twice gives
\begin{equation}
 Y_{\ell-1}=\left(\mathbb E_\ell\mathbb E_{\ell+1}Y_{\ell+1}^{\lambda_\ell}\right)^{1/\lambda_\ell},\qquad \lambda_{\ell+1}=\lambda_\ell>0.
 \label{eq:operational-equal-exponent-composition}
\end{equation}
The independent Gaussian increments enter through their sum, whose variance is the sum of their variances. Their integration therefore coincides with that of the combined interval. At exponent zero, the same statement follows from successive ordinary expectations. The correction term recomposes simultaneously:
\begin{equation}
 \lambda_\ell(q_\ell^2-q_{\ell-1}^2)+\lambda_\ell(q_{\ell+1}^2-q_\ell^2)=\lambda_\ell(q_{\ell+1}^2-q_{\ell-1}^2).
 \label{eq:operational-quadratic-recomposition}
\end{equation}
Subdivision at constant exponent thus preserves both cavity contributions. These subdivisions describe the same effective level, consistently with the treatment of merged layers in Subsection~\ref{subsec:contributi-di-cavita-e-ricorsione-di-parisi}. A geometric partition refined to include the physical overlap thresholds reproduces the physical functional exactly by assigning each subinterval the exponent of the physical level it represents.

Every admissible ordered overlap profile and sequence of mass exponents admits this construction. Conversely, ordered cumulative block fractions and mass exponents specify a hierarchical trial ROSt through the Gaussian construction above. Refining a prescribed geometric mesh allows the overlap thresholds to be approximated, and continuity of the Parisi functional preserves the limiting variational value.

For SK, the parameters can consequently be selected directly in the operational form~\eqref{eq:operational-parisi-functional}, according to the variational principle of Refs.~\cite[Eq.~(6)]{SimplifiedParisi} and~\cite[Sec.~6.1]{AizenmanSimsStarr}. A finite hierarchy gives a variational approximation. Minimization over the ordered fractions and mass exponents, followed by refinement of the hierarchy, recovers the thermodynamic pressure~\cite{GuerraRSBBound,TalagrandParisiFormula}. The physical construction explains how the masses and correlations arise from the reservoir; the operational construction represents those same data through block dimensions and provides the corresponding CCM calculation.

The applications to Ising restricted and deep Boltzmann machines~\cite{CCMBoltzmannMachines} and to the Ising perceptron~\cite{CCMIsingPerceptron} demonstrate that this operational construction extends to different interaction structures: the constructive decomposition determines the cavity fields appropriate to each model, while their identification with the ROSt and recursive averaging provide a common procedure for deriving the corresponding functionals.

We conclude by observing that the derivation above, and in particular Eq.~\eqref{eq:operational-overlap-recomposition}, also identifies the correct interpretation of the state described in Ref.~\cite{ConstructiveCavity} as the limit of vanishing layer magnetizations on the fiber, corresponding to the annealed regime. Thus, the Cartesian product of independent REM measures considered in Ref.~\cite{ConstructiveCavity} admits an annealed realization that clarifies its interpretation. Setting the layer magnetizations to zero in Eq.~\eqref{eq:operational-overlap-recomposition} makes the overlap parameters coincide with the cumulative layer fractions. In the diffuse limit in which all mass exponents approach one, the recursion reduces to ordinary Gaussian averages, and Eq.~\eqref{eq:operational-parisi-functional} gives the annealed pressure independently of the geometric partition. In this limit, the residual factors in Eqs.~(35)--(40) of Ref.~\cite{ConstructiveCavity} concentrate at unity, making their cancellation valid. The Cartesian construction therefore realizes a replica-symmetric annealed case written in the multilevel form of the Parisi functional. The interpretative error concerns identifying this annealed representative and its geometric parameters with the physical Gibbs fiber at low temperature. This distinction is directly related to the use of the Blanket measure without the Gibbs tilt, which also yields the annealed value, as we discuss in the following subsection.

\subsection{The tilt between the Blanket measure and the Gibbs measure}
\label{subsec:il-tilt-fra-misura-blanket-e-misura-gibbs}

In this section we clarify an ambiguity never explicitly addressed in previous works concerning the use of the Blanket measure (with locally normalized weights) in place of the Gibbs measure (globally normalized) in the Aizenman--Sims--Starr cavity formula. In previous works on the CCM, particularly in Ref.~\cite{SimplifiedParisi}, it is used in place of the Gibbs measure in the formula without further specifications. However, the two measures are related by a non-negligible tilt that can generally change the reference physical fiber, and indeed it can be shown that the naive application of the Blanket measure as in Ref.~\cite{SimplifiedParisi} leads to an annealed functional form. In this section we clarify the correct use of the Blanket measure~\cite{BlanketTheory} in the cavity formula and the role of the tilt in changing the physical fiber on which the substitution of the two measures is valid. Here we show how, by applying a further refinement of the partition~\mbox{\cite{RSBWithoutReplicas,BlanketTheory}}, the tilt becomes constant on the fiber, canceling in the cavity formula and thus allowing the substitution of the two measures on the correct physical fiber.

At fixed prefix, the partition function of the rectified interface is
\begin{equation}
 \bar Z_\ell^*(\sigma_{Q_{\ell-1}}^*)=\sum_{\sigma_{V_\ell}^*\in\Omega^{V_\ell}}e^{-\beta\bar H_\ell^*(\sigma_{Q_\ell}^*)}.
 \label{eq:blanket-local-partition-function}
\end{equation}
Linearity in the layer spins allows the sum to be evaluated:
\begin{equation}
 \bar Z_\ell^*(\sigma_{Q_{\ell-1}}^*)=\prod_{i\in V_\ell}2\cosh\!\left[\beta\sqrt{q_{\ell-1}}\,h_i^*(\sigma_{Q_{\ell-1}}^*)\right],\qquad \ell>1.
 \label{eq:blanket-local-cosh-normalizer}
\end{equation}
For the first layer the energy is zero and the partition function counts all its configurations:
\begin{equation}
 \bar Z_1^*(\sigma_{Q_0}^*)=2^{N_1}.
 \label{eq:blanket-first-local-normalizer}
\end{equation}
The locally normalized measure is
\begin{equation}
 \bar\xi_\ell^*(\sigma_{V_\ell}^*\mid\sigma_{Q_{\ell-1}}^*)=\frac{e^{-\beta\bar H_\ell^*(\sigma_{Q_\ell}^*)}}{\bar Z_\ell^*(\sigma_{Q_{\ell-1}}^*)}.
 \label{eq:blanket-local-measure}
\end{equation}
The black dot denotes the Blanket measure; the asterisk retains the calibration on the ground state:
\begin{equation}
 \bar\mu^{\bullet *}(\sigma_V^*)=\prod_{\ell=1}^L\bar\xi_\ell^*(\sigma_{V_\ell}^*\mid\sigma_{Q_{\ell-1}}^*).
 \label{eq:blanket-product-measure}
\end{equation}
This measure sums to one: the spins of the last layer are summed over and the procedure is repeated on the preceding ones. The product thus describes the sequence of local samplings conditioned on the prefixes.

\begingroup\emergencystretch=1em Multiplying again by the local partition functions reconstructs the Boltzmann weight of the entire Hamiltonian. The sum over the configurations gives the identity of Ref.~\cite[Eqs.~(52)--(53)]{BlanketTheory}:\par\endgroup
\begin{equation}
 \bar Z=\left\langle\prod_{\ell=1}^L\bar Z_\ell^*(\sigma_{Q_{\ell-1}}^*)\right\rangle_{\bar\mu^{\bullet *}}.
 \label{eq:blanket-gibbs-normalizer-identity}
\end{equation}
The complete Gibbs measure of~\eqref{eq:rectified-gibbs-measure} is therefore obtained through the tilt
\begin{equation}
 \bar\mu^*(\sigma_V^*)=\bar\mu^{\bullet *}(\sigma_V^*)\frac{\displaystyle\prod_{\ell=1}^L\bar Z_\ell^*(\sigma_{Q_{\ell-1}}^*)}{\displaystyle\left\langle\prod_{\ell=1}^L\bar Z_\ell^*(\sigma_{Q_{\ell-1}}^*)\right\rangle_{\bar\mu^{\bullet *}}}.
 \label{eq:blanket-gibbs-tilt}
\end{equation}
The normalizing factors of the subsequent layers depend on the prefixes already constructed. Their product therefore changes the masses of the continuations. This is the same role played by the totals of the children in~\eqref{eq:two-level-complete-parent-mass}. The common rectification transports both measures and their ratio, preserving this dependence.

Applying~\eqref{eq:blanket-gibbs-tilt} to the two contributions, the global normalization cancels in the difference of the logarithms:
\begin{equation}
 \begin{aligned}
 A(\bar\mu^*)={}&\log2+\log\left\langle\left[\prod_{\ell=1}^L\bar Z_\ell^*(\sigma_{Q_{\ell-1}}^*)\right]\cosh[\beta\tilde x^*(\sigma_V^*)]\right\rangle_{\bar\mu^{\bullet *}}\\
 &-\log\left\langle\left[\prod_{\ell=1}^L\bar Z_\ell^*(\sigma_{Q_{\ell-1}}^*)\right]e^{\beta\tilde y^*(\sigma_V^*)}\right\rangle_{\bar\mu^{\bullet *}}.
 \end{aligned}
 \label{eq:blanket-gibbs-cavity-tilt}
\end{equation}
The product of the normalizing factors remains inside both averages. The recursion on the complete masses already incorporates this weight before evaluating the observables.

The stepwise model of Ref.~\cite[Sec.~3.1, Eqs.~(81)--(87)]{BlanketTheory} allows us to specify when the tilt becomes constant on the fiber. At fixed prefix, the field components are ordered and each layer is subdivided into sublayers on which the field is approximately constant:
\begin{equation}
 V_\ell=\bigcup_{t=1}^L V_{\ell t}.
 \label{eq:blanket-second-partition}
\end{equation}
The sublayer magnetizations are
\begin{equation}
 m_{\ell t}=\frac{M(\sigma_{V_{\ell t}}^*)}{|V_{\ell t}|}.
 \label{eq:blanket-refined-magnetizations}
\end{equation}
The stepwise representative of the interface takes the form
\begin{equation}
 \bar H_\ell^*(\sigma_{Q_\ell}^*)\simeq\sqrt{q_{\ell-1}}\sum_{t=1}^L h_{\ell t}^*(\sigma_{Q_{\ell-1}}^*)M(\sigma_{V_{\ell t}}^*),\qquad\ell>1.
 \label{eq:blanket-step-hamiltonian}
\end{equation}
Its local partition function is therefore
\begin{equation}
 \bar Z_\ell^*(\sigma_{Q_{\ell-1}}^*)\simeq\prod_{t=1}^L\left[2\cosh\!\left(\beta\sqrt{q_{\ell-1}}\,h_{\ell t}^*(\sigma_{Q_{\ell-1}}^*)\right)\right]^{|V_{\ell t}|},\qquad\ell>1.
 \label{eq:blanket-step-local-normalizer}
\end{equation}
In the description by profiles, the step values of the fields, the sublayer volumes, and their magnetizations are fixed together. The representation through magnetizations fixed on groups of sites identified by the fields connects to the cell description of Ref.~\cite{MicrostructureUltrametricity}, already recalled in Ref.~\cite[Sec.~3.1, after Eq.~(85)]{BlanketTheory}. On each fiber thus specified, the right-hand side depends only on these data, while the microscopic positions of the defects can vary. In the stepwise model the product of the normalizing factors is therefore constant on the fiber. Equation~\eqref{eq:blanket-gibbs-tilt} shows that the normalized restrictions of the two measures to this same fiber coincide; the common factor also cancels in the two conditional cavity contributions.

The description through complete Gibbs masses retains the relative weight of the different fibers and continuations. The comparison with the Blanket measure thus identifies the content of the local normalizations: they produce the exact tilt of~\eqref{eq:blanket-gibbs-tilt}, while their constancy on the fixed profile of the stepwise model allows the local calculation directly on the fiber.

Note that in an earlier work on the CCM~\cite{RSBWithoutReplicas} that uses the Blanket measure, the rectification is performed with respect to a direction generally different from the ground state, the \emph{master direction}~\cite{RSBWithoutReplicas}, which depends on the incoming field of the individual layer. It is the reference for the rectification of the normalized local measures that make up the Blanket measure~\cite[Eqs.~(68)--(70)]{BlanketTheory}. In the physical coordinates it is defined by the direction of the incoming field, which depends on the prefix~\cite[Eqs.~(4.25)--(4.29)]{RSBWithoutReplicas}:
\begin{equation}
 \omega_i^\bullet(\sigma_{Q_{\ell-1}})=\frac{h_i(\sigma_{Q_{\ell-1}})}{|h_i(\sigma_{Q_{\ell-1}})|},\qquad i\in V_\ell,\quad\ell>1.
 \label{eq:blanket-master-direction}
\end{equation}
At zero field we choose the positive sign. With the energy convention adopted, at zero temperature the unconstrained local minimum has spins opposite to the field. The associated change of coordinates is
\begin{equation}
 \sigma_{V_\ell}^{\bullet}=\sigma_{V_\ell}\circ\omega_{V_\ell}^{\bullet}(\sigma_{Q_{\ell-1}}).
 \label{eq:blanket-master-coordinate-map}
\end{equation}
The first layer remains unchanged. The map is triangular and its inverse reconstructs the layers in increasing order, each time using the physical prefix already determined. It makes the coefficients of the local linear problem positive and preserves its normalizing factors, by the evenness of the hyperbolic cosine.

The Blanket measure also admits rectification on the ground state through the transformation~\eqref{eq:ground-state-spin-map}, transporting weights and observables together. This is the choice adopted in this subsection: the dot denotes the Blanket measure and the asterisk the calibration on the ground state. On the refined fiber, the constancy of the tilt identifies the normalized restrictions of the two measures. The common calibration on the ground state therefore allows them to be compared in the same coordinates. The global reference is determined by the entire Hamiltonian, also including the interactions with the subsequent layers in~\eqref{eq:single-defect-stability}.

\section{Conclusions}
\label{sec:conclusions}

We have developed a physical justification of the CCM for the SK model, connecting the microscopic description of excitations to the Aizenman--Sims--Starr ROSt of Refs.~\cite{AizenmanSimsStarr,AizenmanSimsStarrOriginal}. The layered decomposition and the dilute gas of defects relative to the ground state allow the complete energies, Gibbs masses, and overlaps to be described on the same configurations. At a fixed kernel, the latter determine the joint law of the complete fresh fields, including the common components. The evaluation of the two cavity contributions in the hierarchical representative yields the Parisi functional. Subsection~\ref{subsec:operational-form-of-the-ccm} translates this identification into the operational CCM: a suitable choice of constructive block dimensions preserves the masses and the joint law of the complete fields, yielding the same cavity functional. The comparison between the Blanket measure and the Gibbs measure also specifies the conditions that allow their normalized restrictions on the refined physical fiber to be substituted for one another.

The sharing of defects and the renewal of fluctuations after the branches separate provide a physical motivation for the GREM structure and the ultrametricity of the kernel, developed from the case with finitely many defects to the regime with exponential populations. The representation through independent PPPs, ancestor assignment, and global normalization reconstructs the ultrametric cascade and corrects its identification with a Cartesian product of independent REM measures in Ref.~\cite{ConstructiveCavity}. The joint transport of masses and fields preserves the random functional of the cascade in distribution; the calculation of the mean logarithms corrects the earlier passage to the Parisi recursion.

\subsection{Physical justification of the CCM--ROSt identification}
\label{subsec:giustificazione-fisica-dell-identificazione-ccmrost}

The main result is the joint identification of the two cavity contributions with those of the same Aizenman--Sims--Starr ROSt~\cite{AizenmanSimsStarr}, formulated in~\eqref{eq:rost-linear-insertion-identification} and~\eqref{eq:rost-quadratic-insertion-identification}. The kernel connects the Gibbs masses, determined by the complete energies of the branches, to the overlaps of the same configurations. States that share a prefix inherit the same defects and the same energy contributions; after the separation, the new nodes add new arrangements. This mechanism motivates the GREM representation developed in Section~\ref{sec:grem-universality} and provides the physical justification for connecting the constructive variables of Ref.~\cite{SimplifiedParisi} to the ROSt. The two contributions are thus evaluated on the same states and with the same weights, preserving the dependence required by the random cavity functional.

The identification of the fields follows from an exact Gaussian calculation at a fixed kernel. The covariances~\eqref{eq:full-linear-cavity-covariance} and~\eqref{eq:full-quadratic-cavity-covariance} are, respectively, twice the overlap and its square. Rectification relative to the ground state preserves the overlaps and, since the reference depends only on the reservoir disorder, leaves the joint law of the fresh noises and their independence from the energy weights unchanged. The joint representation of the masses and overlaps therefore determines that of the two families of fields. At a fixed temperature, control of the omitted mass and of the uniform Gaussian moments allows the connection to be extended from a finite number of states to the complete sums and to the averages of the functional, as discussed in Subsection~\ref{subsec:ansatz-di-convergenza-alla-rost}. The recombination of the fibers according to~\eqref{eq:gibbs-fibre-weight} and~\eqref{eq:observable-disintegration} preserves their Gibbs weights and the same fresh noises, maintaining the joint character of the identification.

The ROSt amplitudes have an explicit geometric meaning. The volume fractions and the layer magnetizations determine the overlap increments through~\eqref{eq:hierarchical-overlap-increment}; these yield the amplitudes of the linear field and, through the differences of the squares of the cumulative overlaps in~\eqref{eq:hierarchical-quadratic-increment}, those of the correction field. The mass exponents are determined by the state populations and by the slopes of the complete energy law after rectification relative to the ground state, through~\eqref{eq:grem-complete-energy-rate}, \eqref{eq:grem-general-group-slope}, and~\eqref{eq:grem-group-mass-exponent}. The rescaling~\eqref{eq:grem-group-rem-size}--\eqref{eq:grem-group-rem-partition}, developed starting from Ref.~\cite[Eqs.~(133)--(137)]{REMUniversality}, expresses the effective REM while preserving the cardinality and energy amplitude of the group. In the cavity functional, the fresh field amplitudes remain fixed by the overlaps and both contributions carry the physical temperature of the system.

The treatment of the common components specifies how the fluctuations around the reference considered in Ref.~\cite[Lemma~13]{RSBWithoutReplicas} enter the complete cavity formula. Orthogonality on the fiber makes each common component independent of its respective family of centered fluctuations. In the linear field, the common component acts as a Gaussian field on the new spin: it remains inside the hyperbolic cosine and stays fixed during the recursive averages over subsequent levels. Its final integration corresponds to the initial step with exponent zero of the Parisi functional. In the correction field, the common component separates from the logarithm of the exponential contribution; it is present in the random functional and has zero mean over the fresh noise. Equations~\eqref{eq:rost-complete-linear-field} and~\eqref{eq:rost-complete-quadratic-field} bring these components and the hierarchical increments together in the complete fields of the ROSt.

\begingroup
\emergencystretch=1em
In the hierarchical representative, the equivalence~\eqref{eq:marked-input-cavity-functional} preserves the random functional in distribution. Its mean evaluation gives~\eqref{eq:physical-overlap-parisi-functional}: the logarithmic identity~\eqref{eq:normalized-poisson-mean-log} produces the power means of the recursion~\eqref{eq:parisi-power-recursion}, while the Gaussian moment~\eqref{eq:quadratic-increment-gaussian-average} determines the quadratic correction. The Parisi functional is thus obtained with the physical overlaps and with the normalization of Ref.~\cite[Eqs.~(6)--(8)]{SimplifiedParisi}. Layers with the same exponent are recombined into a single effective level, and the thermal continuations are integrated through the ordinary average. The calculation therefore preserves the contribution of the different parts of the hierarchy in the same formula.
\par\endgroup

Subsection~\ref{subsec:operational-form-of-the-ccm} establishes the correspondence between the physical representation and the operational calculation. Equation~\eqref{eq:operational-overlap-recomposition} collects the alignment shared by all states into the common component and assigns the remaining correlations to the fluctuations inherited along the prefix. The fractions~\eqref{eq:operational-partition-fractions} represent these contributions through the block dimensions of a chosen constructive partition. The state masses and ancestry are retained, and the complete Gaussian fields have the same joint law. The two representations therefore describe the same statistical response of the reservoir in the cavity formula and preserve the distribution of its random functional. The initial common block is integrated at exponent zero, giving the operational form~\eqref{eq:operational-parisi-functional} at the physical temperature of the system.

Subdivision of these blocks at constant mass exponent preserves both cavity contributions: the successive Gaussian averages compose according to~\eqref{eq:operational-equal-exponent-composition}, while the quadratic correction recomposes through~\eqref{eq:operational-quadratic-recomposition}. Each ordered physical overlap profile can thus be represented by suitable block dimensions and their refinements. This correspondence allows the dimensions of the constructive blocks and the mass exponents to be used directly as parameters in the same variational calculation.

Minimization of the functional selects the ordered overlap values and the mass exponents, according to the variational principle of Refs.~\cite[Eq.~(6)]{SimplifiedParisi} and~\cite[Sec.~6.1]{AizenmanSimsStarr}. The volume fractions and magnetizations provide their geometric representation; the populations and energy laws describe their balance between energy and entropy. A finite number of levels corresponds to an approximate variational description, whose subsequent refinement yields the thermodynamic pressure of SK. Control of the cores in~\eqref{eq:interface-free-energy} in turn connects the pressure of the interface-only model to that of the complete model, in the order of limits established in Subsection~\ref{subsec:decomposizione-blanket-e-dominanza-dell-interfaccia}.

The final comparison between the Blanket measure~\cite{BlanketTheory} and the Gibbs measure finally clarifies the residual ambiguities in previous works~\cite{SimplifiedParisi,RSBWithoutReplicas} by specifying the correct use of this measure in the cavity formula and which weights enter the construction. The tilt~\eqref{eq:blanket-gibbs-tilt}, given by the product of the local normalizing factors divided by its Blanket average, modifies the continuation masses and the relative weight of the fibers. Formula~\eqref{eq:blanket-gibbs-cavity-tilt} preserves its effect in both cavity contributions. On the joint profile of fields, volumes, and magnetizations of the stepwise model of Ref.~\cite[Sec.~3.1]{BlanketTheory}, the product is constant: the normalized restrictions of the two measures coincide and the factor cancels in the conditioned contributions. The recombination maintains the Gibbs weights of the different fibers. This clarifies the use of the Blanket measure in the cavity formula of previous works~\cite{SimplifiedParisi,RSBWithoutReplicas,BlanketTheory}; the common rectification relative to the ground state allows the comparison to be made in the same coordinates used for the identification with the ROSt.

\subsection{Ultrametricity in the kernel representation}
\label{subsec:ultrametricita-nella-rappresentazione-mediante-kernel}

The kernel description~\cite{RSBWithoutReplicas} connects the hierarchy of configurations to the composition of their energy contributions. States that share a prefix have the same spins and the same defects in the corresponding layers; when the branch splits, the new nodes receive new arrangements. The last common ancestor thus determines both shared parts. The structure depicted in Ref.~\cite[Figs.~5.2--5.3]{RSBWithoutReplicas} and represented here with defects in Figure~\ref{fig:kernel-grem-gerarchia} thereby acquires a physical motivation through the dilution of defects, developed here while retaining the complete energy in the ground state coordinates.

With a finite number of defects and nodes, uniform sampling produces disjoint supports with probability tending to one over the entire kernel strip, even between nodes with different parents. Interactions between defects give a vanishing contribution and~\eqref{eq:finite-node-energy-increment} expresses the increments through the complete costs of the layer. Sampling this population jointly approximates the new increments by independent sums, while descendants reuse exactly those of their ancestors. Equation~\eqref{eq:finite-tree-overlap} describes the corresponding hierarchy of distances: it depends on the last shared level, before the normalized overlaps all tend to one in the limit with finitely many defects. The calculation therefore makes the physical mechanism of the GREM structure explicit.

The passage to the extensive regime preserves this organization on exponential populations. The thermodynamic limit is first taken at a fixed number of layers, volume fractions, and positive defect densities, with the dilute regime considered subsequently. The preceding strips group the terminal states through the sharing of spins, with the increasing node count of~\eqref{eq:hierarchical-node-count}. The shell capacity, expressed by~\eqref{eq:layer-state-capacity}, and the control of intersections in~\eqref{eq:extensive-population-condition} allow exponential populations to be chosen simultaneously in all layers. The correction~\eqref{eq:extensive-overlap-correction} remains small relative to the distance due to the defects and the control includes all pairs of nodes in the strip. This specifies the relationship between defect densities, admissible populations, and the accuracy of the ultrametric geometry.

The recombination~\eqref{eq:centered-prefix-energy-recomposition} retains all interactions in the energy: each increment includes the field of the prefix already chosen and the mean contribution of subsequent layers. The center~\eqref{eq:physical-fibre-energy-center}, constant on the fiber, accompanies the energy comparison between profiles. At fixed prefixes, sampling of the new nodes factorizes and the energy increments have zero conditional mean; the covariance of the terminal energies therefore collects the contributions of the common ancestors. Renewal of the samples and dilution motivate the passage from the laws conditioned on the individual prefixes to a common law for nodes at the same level. This passage constitutes the physical GREM approximation in the extensive regime.

The local energy law is determined on the actual population of rectified field coefficients, by sampling supports with the prescribed number of defects. The regularity required for counting rare energies is motivated in Subsection~\ref{subsec:passaggio-al-regime-estensivo} through simultaneous control of the possible choices of the rectification and prefix signs. This control preserves a positive fraction of sufficiently dispersed coefficients even for the choice selected by the ground state; control of the matrix norm also bounds their spatial second moment. The connection with the REM mechanism of Ref.~\cite{REMUniversality} therefore uses the complete law of the selected field and sampling on the shell, within the framework of universality for linear energies developed in Refs.~\cite{REMLinear,REMLinearLetter}.

The balance between energy and entropy then determines which layers form distinct effective levels. The cost function~\eqref{eq:grem-complete-energy-rate} describes the rarity of energies through the complete law of the increments. The constraint~\eqref{eq:grem-prefix-energy-cost} requires that the number of paths remain available at each prefix; the conditions~\eqref{eq:grem-general-group-edge}--\eqref{eq:grem-general-group-slope} identify the energy edge of a group and its common slope. Grouping consecutive layers according to the minimum critical inverse temperature produces increasing thresholds along the hierarchy, collecting the layers that condense together into the same effective level. This yields the energy and entropy criterion for the effective levels~\cite{BovierKurkova,JanaRaoLDP}.

The connection between this balance and the weights of the measure proceeds through counting over finite energy differences. Treating the energy and the number of defects jointly determines the probability prefactor on the shell and the resulting logarithmic correction to the centering, according to the tail evaluation mechanism discussed in Ref.~\cite{BovierMayer}. The edge is calculated on the actual population of coefficients, retaining the dependence on the disorder and the fiber profile. For groups, the constraints on the internal prefixes are also retained. The edge slope then determines the exponential intensity~\eqref{eq:grem-extreme-energy-intensity} of the extreme energies. In the GREM representative, independent renewal of the draws produces the Poisson counts of rare energies. The energy factors that depend on the prefix remain in the complete masses.

In the condensed regime, the decay of the Boltzmann weights prevails over the growth of the number of states. Control of the contribution of energies above a window and of the probability of finding states far below it allows the limit of the extremes to be extended to the total and the normalized weights. The ratio between the critical slope and the inverse of the physical temperature determines the exponents in~\eqref{eq:grem-group-mass-exponent}, ordered according to~\eqref{eq:ordered-grem-mass-exponents}. The condensed groups occupy the initial part of the hierarchy; any final thermal groups contribute through the ordinary average.

The representation thus brings together, on the same kernel, the spin geometry, the complete energy, and the entropy of the populations. The sharing of defects determines the overlaps, while the energy and entropy balance determines the effective levels and their weights. These two results provide, respectively, the geometric coefficients and the structure of the measure used in the identification with the ROSt in Section~\ref{sec:cavity-rost}.

The comparison with the droplet description thus acquires geometric and energetic content: the shared defect supports organize the collective excitations, while the complete energies and branch weights describe their thermodynamic relevance. A natural application of the GREM representation is the study of the minimum costs of excitations at a fixed overlap. The analysis of cascades at zero temperature in Ref.~\cite{FranzParisiZeroTemperature} provides a tool for this comparison, considering the minima over the descendant configurations of branches that separate from the ground state at the different levels. A further development concerns the response to small changes in the external field and the resulting equilibrium magnetization jumps, studied for SK in Ref.~\cite{LeDoussalMullerWieseAvalanches}. These applications would allow the organization of excitations constructed through kernels to be compared with observables beyond the free energy.

\subsection{Meaning of the PPP product in the CCM}
\label{subsec:significato-del-prodotto-di-ppp-nel-ccm}

Subsections~\ref{subsec:cascata-grem-e-prodotto-di-ppp-a-due-livelli} and~\ref{subsec:prodotto-di-ppp-ed-equivalenza-in-distribuzione-con-la-cascata} correct the identification, proposed in Ref.~\cite[Eqs.~(18)--(24)]{ConstructiveCavity}, between a Cartesian product of independent REM measures and the cascade used to calculate the Parisi functional. The error arises from the interpretation of the triangular map satisfied by the IO Hamiltonian: the factorization of the new draws is conditional on the preceding levels, while the increments retain their dependence on the prefixes and are shared by the descendants. The construction developed here specifies which processes can be chosen independently and how the masses, ancestors, and observables of the cascade are reconstructed from them.

Global normalization determines the weight of the configurations and their prefixes. Already at two levels, the complete parent mass in~\eqref{eq:two-level-complete-parent-mass} includes the total of the continuations: this factor weights the prefix before normalization over the set of states. The calculation therefore keeps the points of the local processes, their totals, and the normalized weights together, according to the cascade structure described in Ref.~\cite[Sec.~5.2]{AizenmanSimsStarr}. The total and weights of each process derive from the same points and preserve their joint law. Independence concerns the processes used at the different levels.

The mechanism of the product representation is explicit in the two-level calculation. With the first process fixed, superposition of the lists of children produces a single PPP, whose intensity is the sum of the local intensities. The factor~\eqref{eq:two-level-common-mass-scale} allows it to be generated through a process independent of the first; the probabilities~\eqref{eq:two-level-parent-assignment} then assign a parent to each point and reconstruct the original lists. These probabilities determine ancestry, while the sum of the final weights of the descendants determines the physical mass of each parent. The total, the complete masses, and the ancestor structure are thus recovered jointly, through the superposition and subdivision operations underlying the coagulation of Poisson--Dirichlet weights~\cite[Theorem~12]{PitmanCoagulation}.

Iteration extends this construction to the entire hierarchy. The powers of the points in~\eqref{eq:powered-input-processes} produce processes whose intensities are governed by the ratios between consecutive exponents. The assignments~\eqref{eq:multilevel-parent-assignment} reconstruct the complete addresses: each node receives a parent only once and all its descendants inherit the same choice. The identity~\eqref{eq:multilevel-total-product} preserves the cascade total in distribution, while~\eqref{eq:independent-terminal-mass}--\eqref{eq:independent-complete-prefix-mass} reconstruct its terminal masses and its prefix masses. The inverse description, obtained by subdividing the complete masses among the children, follows the fragmentation laws~\eqref{eq:complete-prefix-fragmentation} of Ref.~\cite[Corollary~13]{PitmanCoagulation}. The total, masses, and ancestry therefore jointly have the law of the RPC.

\begingroup
\emergencystretch=1em
The product therefore concerns the law of the independent input processes. The measure resulting from ancestor assignment and global normalization has the ultrametric geometry of the cascade, including in the spin overlaps. The configurations associated with descendants indeed share the spins and defects of the common prefixes. This construction corrects the statement in Ref.~\cite{ConstructiveCavity} about the possibility of obtaining the Parisi functional through the non-ultrametric Cartesian product. The use of a single index to enumerate the nodes at each level is compatible with the hierarchy because the ancestors remain determined by the assignment map.

The correspondence also includes the observables. The same assignments transport the configurations, the masses, and the two families of cavity fields; descendants reuse the Gaussian increments of their ancestors. The correlations between states required for the calculation are thus preserved, specifying the meaning of suppressing the prefixes in Eq.~(30) of Ref.~\cite{ConstructiveCavity}. Equation~\eqref{eq:marked-input-cavity-functional} expresses the equivalence in distribution of the random functional of the hierarchical representation, with both contributions evaluated on the same masses and the same ancestry. The common global normalization cancels algebraically in the ratio.

\par\endgroup

The evaluation of the mean also corrects the passage in Eqs.~(35)--(40) of Ref.~\cite{ConstructiveCavity}. In that calculation, the distributional identities of the individual marked sums are used to cancel residual factors between the two contributions. The marginal laws, however, leave undetermined the dependence between these factors, on which their cancellation depends. The calculation in Subsection~\ref{subsec:contributi-di-cavita-e-ricorsione-di-parisi} instead uses the law of the marked sum~\eqref{eq:marked-poisson-total-law} to calculate its mean logarithm and subtracts the mean logarithm of the original total. Linearity of the mean yields the identity~\eqref{eq:normalized-poisson-mean-log}, which determines the power mean required at each level.

The two-level calculation~\eqref{eq:two-level-parisi-average} shows how the ratios between the exponents of the transformed processes yield, through composition of the averages, the exponents of the cascade levels. Iteration from the last level to the first produces the recursion~\eqref{eq:parisi-power-recursion} and, by bringing the two cavity contributions together, the Parisi functional~\eqref{eq:physical-overlap-parisi-functional}. The procedure applies to the effective condensed groups, preserving the ordinary averages of any thermal layers in the continuations. This yields an operational formulation through independent PPPs and ancestor assignment: the construction preserves the random functional of the cascade in distribution and its recursive average yields the Parisi functional.

\section*{Acknowledgments}\label{sec:acknowledgments}
We thank Francesco Concetti (Unidistance), Giampiero Bardella (Sapienza Universita di Roma), Nicola Kistler (Goethe University Frankfurt) and ChatGPT (OpenAI) for important discussions and suggestions.

\section*{Declaration of Generative AI}
Research, writing, and proofreading of the present manuscript were performed with AI assistance (GPT-6 Astra) under strict human guidance and scrutiny. We independently checked all mathematical arguments and take full responsibility for the content of the publication.


\begin{thebibliography}{99}
\addcontentsline{toc}{section}{\refname}
\small\setstretch{1.05}\raggedright

\bibitem{SimplifiedParisi}
S. Franchini, \emph{A simplified Parisi Ansatz},
Communications in Theoretical Physics \textbf{73}, 055601 (2021).
\href{https://doi.org/10.1088/1572-9494/abde32}{doi:10.1088/1572-9494/abde32}.
Cited preprint version:
\href{https://arxiv.org/abs/1909.06594v7}{arXiv:1909.06594v7}.

\bibitem{RSBWithoutReplicas}
S. Franchini, \emph{Replica Symmetry Breaking without Replicas},
Annals of Physics \textbf{450}, 169220 (2023).
\href{https://doi.org/10.1016/j.aop.2023.169220}{doi:10.1016/j.aop.2023.169220}.
Cited preprint version:
\href{https://arxiv.org/abs/1610.03941v13}{arXiv:1610.03941v13}.

\bibitem{SherringtonKirkpatrick}
D. Sherrington and S. Kirkpatrick, \emph{Solvable Model of a Spin-Glass},
Physical Review Letters \textbf{35}, 1792--1796 (1975).
\href{https://doi.org/10.1103/PhysRevLett.35.1792}{doi:10.1103/PhysRevLett.35.1792}.

\bibitem{BlanketTheory}
S. Franchini, \emph{Universal scaling limits for spin networks via martingale methods},
Proceedings of the Royal Society A \textbf{481}, 20240774 (2025).
\href{https://doi.org/10.1098/rspa.2024.0774}{doi:10.1098/rspa.2024.0774}.
Cited preprint version:
\href{https://arxiv.org/abs/2410.08150v5}{arXiv:2410.08150v5}.

\bibitem{Multispin}
S. Franchini, \emph{Blanket representation for spin networks with
$n$-body interactions},
\href{https://arxiv.org/abs/2503.21101v3}{arXiv:2503.21101v3} (2025).
\href{https://doi.org/10.48550/arXiv.2503.21101}{doi:10.48550/arXiv.2503.21101}.

\bibitem{AizenmanSimsStarrOriginal}
M. Aizenman, R. Sims and S. L. Starr, \emph{Extended variational principle for the Sherrington--Kirkpatrick spin-glass model},
Physical Review B \textbf{68}, 214403 (2003).
\href{https://doi.org/10.1103/PhysRevB.68.214403}{doi:10.1103/PhysRevB.68.214403}.

\bibitem{AizenmanSimsStarr}
M. Aizenman, R. Sims and S. L. Starr, \emph{Mean-field spin glass models
from the cavity--ROSt perspective},
Contemporary Mathematics \textbf{437}, 1--30 (2007).
\href{https://doi.org/10.1090/conm/437/08422}{doi:10.1090/conm/437/08422}.
Cited preprint version:
\href{https://arxiv.org/abs/math-ph/0607060v1}{arXiv:math-ph/0607060v1}.

\bibitem{ConstructiveCavity}
S. Franchini, \emph{Constructive Cavity Method},
Journal of Mathematical Physics \textbf{67}, 013302 (2026).
\href{https://doi.org/10.1063/5.0291580}{doi:10.1063/5.0291580}.
Cited preprint version:
\href{https://arxiv.org/abs/2601.00410v1}{arXiv:2601.00410v1}.

\bibitem{DerridaREM}
B. Derrida, \emph{Random-energy model: An exactly solvable model of disordered systems},
Physical Review B \textbf{24}, 2613--2626 (1981).
\href{https://doi.org/10.1103/PhysRevB.24.2613}{doi:10.1103/PhysRevB.24.2613}.

\bibitem{REMUniversality}
S. Franchini, \emph{A simplified Parisi Ansatz II: Random Energy Model universality},
Chaos, Solitons \& Fractals \textbf{191}, 115821 (2025).
\href{https://doi.org/10.1016/j.chaos.2024.115821}{doi:10.1016/j.chaos.2024.115821}.
Corrigendum: \textbf{201}, 117395 (2025).
\href{https://doi.org/10.1016/j.chaos.2025.117395}{doi:10.1016/j.chaos.2025.117395}.
Cited corrected preprint:
\href{https://arxiv.org/abs/2312.07808v7}{arXiv:2312.07808v7}.

\bibitem{CCMBoltzmannMachines}
S. Franchini, F. Concetti, G. Bardella and L. Pan,
\emph{Constructive Cavity Method for Ising Restricted and Deep Boltzmann Machines with Gaussian Couplings},
\href{https://arxiv.org/abs/2610.36505}{arXiv:2610.36505} (2026).
\href{https://doi.org/10.48550/arXiv.2610.36505}{doi:10.48550/arXiv.2610.36505}.

\bibitem{CCMIsingPerceptron}
S. Franchini, F. Concetti and G. Bardella,
\emph{The Constructive Cavity Method for the Ising Perceptron},
\href{https://arxiv.org/abs/2610.39141}{arXiv:2610.39141} (2026).
\href{https://doi.org/10.48550/arXiv.2610.39141}{doi:10.48550/arXiv.2610.39141}.

\bibitem{MezardParisiVirasoroCavity}
M. M\'ezard, G. Parisi and M. A. Virasoro, \emph{SK Model: The Replica Solution without Replicas},
Europhysics Letters \textbf{1}(2), 77--82 (1986).
\href{https://doi.org/10.1209/0295-5075/1/2/006}{doi:10.1209/0295-5075/1/2/006}.

\bibitem{FisherHuseDroplet}
D. S. Fisher and D. A. Huse, \emph{Equilibrium behavior of the spin-glass
ordered phase}, Physical Review B \textbf{38}, 386--411 (1988).
\href{https://doi.org/10.1103/PhysRevB.38.386}{doi:10.1103/PhysRevB.38.386}.

\bibitem{MertensRandomCosts}
S. Mertens, \emph{Random Costs in Combinatorial Optimization},
Physical Review Letters \textbf{84}, 1347--1350 (2000).
\href{https://doi.org/10.1103/PhysRevLett.84.1347}{doi:10.1103/PhysRevLett.84.1347}.

\bibitem{BaukeFranzMertens}
H. Bauke, S. Franz and S. Mertens, \emph{Number partitioning as a random
energy model}, Journal of Statistical Mechanics: Theory and Experiment
\textbf{2004}, P04003 (2004).
\href{https://doi.org/10.1088/1742-5468/2004/04/P04003}{doi:10.1088/1742-5468/2004/04/P04003}.

\bibitem{BenArousGayrardKuptsov}
G. Ben Arous, V. Gayrard and A. Kuptsov, \emph{A new REM conjecture},
in \emph{In and Out of Equilibrium 2}, Progress in Probability
\textbf{60}, Birkh\"auser, Basel, 59--96 (2008).
\href{https://doi.org/10.1007/978-3-7643-8786-0_4}{doi:10.1007/978-3-7643-8786-0\_4}.

\bibitem{BovierKurkovaLocalREM}
A. Bovier and I. Kurkova, \emph{Local energy statistics in disordered systems:
a proof of the local REM conjecture},
Communications in Mathematical Physics \textbf{263}, 513--533 (2006).
\href{https://doi.org/10.1007/s00220-005-1516-1}{doi:10.1007/s00220-005-1516-1}.

\bibitem{BorgsChayesMertensNairI}
C. Borgs, J. Chayes, S. Mertens and C. Nair, \emph{Proof of the local REM
conjecture for number partitioning. I. Constant energy scales},
Random Structures \& Algorithms \textbf{34}(2), 217--240 (2009).
\href{https://doi.org/10.1002/rsa.20255}{doi:10.1002/rsa.20255}.

\bibitem{BorgsChayesMertensNairII}
C. Borgs, J. Chayes, S. Mertens and C. Nair, \emph{Proof of the local REM
conjecture for number partitioning. II. Growing energy scales},
Random Structures \& Algorithms \textbf{34}(2), 241--284 (2009).
\href{https://doi.org/10.1002/rsa.20256}{doi:10.1002/rsa.20256}.

\bibitem{REMLinear}
F. Concetti and S. Franchini, \emph{REM universality for linear random energy},
\href{https://arxiv.org/abs/2604.06122v1}{arXiv:2604.06122v1} (2026).
\href{https://doi.org/10.48550/arXiv.2604.06122}{doi:10.48550/arXiv.2604.06122}.

\bibitem{REMLinearLetter}
F. Concetti and S. Franchini, \emph{REM universality and Poisson--Dirichlet
Gibbs weights for linear random energy},
\href{https://arxiv.org/abs/2606.07757v2}{arXiv:2606.07757v2} (2026).
\href{https://doi.org/10.48550/arXiv.2606.07757}{doi:10.48550/arXiv.2606.07757}.

\bibitem{DerridaGREM}
B. Derrida, \emph{A generalization of the random energy model which includes correlations between energies},
Journal de Physique Lettres \textbf{46}(9), L401--L407 (1985).
\href{https://doi.org/10.1051/jphyslet:01985004609040100}{doi:10.1051/jphyslet:01985004609040100}.

\bibitem{BovierKurkova}
A. Bovier and I. Kurkova, \emph{Derrida's generalised random energy
models 1: models with finitely many hierarchies},
Annales de l'Institut Henri Poincar\'e, Probabilit\'es et Statistiques
\textbf{40}(4), 439--480 (2004).
\href{https://doi.org/10.1016/j.anihpb.2003.09.002}{doi:10.1016/j.anihpb.2003.09.002}.

\bibitem{JanaRaoLDP}
N. K. Jana and B. V. Rao, \emph{Large Deviations and Random Energy Models},
\href{https://arxiv.org/abs/math/0612836v1}{arXiv:math/0612836v1} (2006).
\href{https://doi.org/10.48550/arXiv.math/0612836}{doi:10.48550/arXiv.math/0612836}.

\bibitem{RuelleCascades}
D. Ruelle, \emph{A mathematical reformulation of Derrida's REM and GREM},
Communications in Mathematical Physics \textbf{108}, 225--239 (1987).
\href{https://doi.org/10.1007/BF01210613}{doi:10.1007/BF01210613}.

\bibitem{PitmanCoagulation}
J. Pitman, \emph{Coalescents with multiple collisions},
The Annals of Probability \textbf{27}(4), 1870--1902 (1999).
\href{https://doi.org/10.1214/aop/1022874819}{doi:10.1214/aop/1022874819}.
Cited technical-report version (revised 23 March 1999):
\href{https://statistics.berkeley.edu/sites/default/files/tech-reports/495.pdf}
{UC Berkeley, Technical Report 495}, Theorem~12 and Corollary~13.

\bibitem{LeDoussalMullerWieseAvalanches}
P. Le Doussal, M. M\"uller and K. J. Wiese, \emph{Equilibrium avalanches
in spin glasses}, Physical Review B \textbf{85}, 214402 (2012).
\href{https://doi.org/10.1103/PhysRevB.85.214402}{doi:10.1103/PhysRevB.85.214402}.

\bibitem{PankovLowTemperature}
S. Pankov, \emph{Low-Temperature Solution of the Sherrington--Kirkpatrick Model},
Physical Review Letters \textbf{96}, 197204 (2006).
\href{https://doi.org/10.1103/PhysRevLett.96.197204}{doi:10.1103/PhysRevLett.96.197204}.

\bibitem{BolthausenKistlerUltrametricity}
E. Bolthausen and N. Kistler, \emph{On a nonhierarchical version of the Generalized Random Energy Model, II: Ultrametricity},
Stochastic Processes and their Applications \textbf{119}(7), 2357--2386 (2009).
\href{https://doi.org/10.1016/j.spa.2008.12.002}{doi:10.1016/j.spa.2008.12.002}.

\bibitem{DerridaGardnerGREM}
B. Derrida and E. Gardner, \emph{Solution of the generalised random energy model},
Journal of Physics C: Solid State Physics \textbf{19}, 2253--2274 (1986).
\href{https://doi.org/10.1088/0022-3719/19/13/015}{doi:10.1088/0022-3719/19/13/015}.

\bibitem{BovierMayer}
A. Bovier and H. Mayer, \emph{A conditional strong large deviation result
and a functional central limit theorem for the rate function},
ALEA, Latin American Journal of Probability and Mathematical Statistics
\textbf{12}(1), 533--550 (2015).
\href{https://alea.impa.br/articles/v12/12-21.pdf}{Published article}.
Preprint DOI: \href{https://doi.org/10.48550/arXiv.1411.5803}{doi:10.48550/arXiv.1411.5803}.

\bibitem{BolthausenSznitmanCascades}
E. Bolthausen and A.-S. Sznitman, \emph{On Ruelle's Probability Cascades and an Abstract Cavity Method},
Communications in Mathematical Physics \textbf{197}, 247--276 (1998).
\href{https://doi.org/10.1007/s002200050450}{doi:10.1007/s002200050450}.

\bibitem{PitmanYorPoissonDirichlet}
J. Pitman and M. Yor, \emph{The two-parameter Poisson--Dirichlet distribution derived from a stable subordinator},
The Annals of Probability \textbf{25}(2), 855--900 (1997).
\href{https://doi.org/10.1214/aop/1024404422}{doi:10.1214/aop/1024404422}.

\bibitem{GuerraToninelliLimit}
F. Guerra and F. L. Toninelli, \emph{The thermodynamic limit in mean field
spin glass models}, Communications in Mathematical Physics
\textbf{230}, 71--79 (2002).
\href{https://doi.org/10.1007/s00220-002-0699-y}{doi:10.1007/s00220-002-0699-y}.
Cited preprint version:
\href{https://arxiv.org/abs/cond-mat/0204280v1}{arXiv:cond-mat/0204280v1}.

\bibitem{MicrostructureUltrametricity}
M. M\'ezard and M. A. Virasoro, \emph{The microstructure of ultrametricity},
Journal de Physique \textbf{46}(8), 1293--1307 (1985).
\href{https://doi.org/10.1051/jphys:019850046080129300}{doi:10.1051/jphys:019850046080129300}.

\bibitem{ParisiOrderParameter}
G. Parisi, \emph{The order parameter for spin glasses: a function on the interval 0--1},
Journal of Physics A: Mathematical and General \textbf{13}, 1101--1112 (1980).
\href{https://doi.org/10.1088/0305-4470/13/3/042}{doi:10.1088/0305-4470/13/3/042}.

\bibitem{GuerraRSBBound}
F. Guerra, \emph{Broken Replica Symmetry Bounds in the Mean Field Spin Glass Model},
Communications in Mathematical Physics \textbf{233}, 1--12 (2003).
\href{https://doi.org/10.1007/s00220-002-0773-5}{doi:10.1007/s00220-002-0773-5}.

\bibitem{TalagrandParisiFormula}
M. Talagrand, \emph{The Parisi formula},
Annals of Mathematics \textbf{163}, 221--263 (2006).
\href{https://doi.org/10.4007/annals.2006.163.221}{doi:10.4007/annals.2006.163.221}.

\bibitem{FranzParisiZeroTemperature}
S. Franz and G. Parisi, \emph{Non trivial overlap distributions at zero
temperature}, The European Physical Journal B \textbf{18}, 485--491 (2000).
\href{https://doi.org/10.1007/s100510070037}{doi:10.1007/s100510070037}.

\end{thebibliography}
\end{document}